\documentclass{aa}  
\usepackage{natbib}
\usepackage{xcolor}
\usepackage{graphicx}
\usepackage{txfonts}
\usepackage{hyperref}
\usepackage{float}
\usepackage{tablefootnote}
\usepackage{lipsum}
\usepackage{caption}
\usepackage{afterpage}
\usepackage{subcaption}
\usepackage{multirow}
\usepackage{svg}
\usepackage{lipsum}
\usepackage{lscape}

\begin{document}

   \title{Cluster-lensed supernova yields from the \textit{Vera C. Rubin} Observatory and \textit{Nancy Grace Roman} Space Telescope}

   \author{M. Bronikowski \inst{1}
          \and
          T. Petrushevska \inst{1}
          \and
          J. D. R. Pierel \inst{2}
          \and
        A. Acebron \inst{3, 4}
          \and
          D. Donevski \inst{5}
          \and
            B. Apostolova \inst{1}
          \and
          N. Blagorodnova \inst{6,7,8}
            \and
          T. Jankovi\v{c} \inst{1,9}
          }

   \institute{Center for Astrophysics and Cosmology, University of Nova Gorica, Vipavska 11c, 5270 Ajdov\v{s}\v{c}ina, Slovenia\\
              \email{mateusz.bronikowski@ung.si}
         \and
            Space Telescope Science Institute, 3700 San Martin Drive, Baltimore, MD 21218, USA
         \and
         Instituto de Física de Cantabria (CSIC-UC), Avda. Los Castros s/n, 39005 Santander, Spain
         \and
         Dpto. de Física Moderna, Universidad de Cantabria, Avda. de los Castros s/n, E-39005 Santander, Spain
        \and         National Center for Nuclear Research, Pasteura 7, 02-093 Warsaw, Poland
        \and Institut de Ciències del Cosmos (ICCUB), Universitat de Barcelona, Gran via de les corts catalanes 585, 08007 Barcelona, Spain
        \and Departament de Física Quántica i Astrofísica (FQA), Universitat de Barcelona (UB), Barcelona, Spain
        \and 
        Institut d'Estudis Espacials de Catalunya (IEEC), Barcelona, Spain
        \and  Institute of Physics of the Czech Academy of Sciences, Na Slovance 1999/2, 182 21 Praha 8, Prague, Czech Republic \\
             }

   \date{Received XX; accepted XX}

  \abstract
   {Through gravitational lensing, galaxy clusters can magnify supernovae (SNe) and thereby create multiple images of the same SN. This enables measurements of cosmological parameters (primarily the Hubble constant), which will be increasingly important in the context of  upcoming surveys  from the \textit{Nancy Grace Roman} Space Telescope (\textit{Roman}) and  \textit{Vera C. Rubin} Observatory.}
   {We study the prospects of detecting strongly lensed supernovae in cluster fiels with \textit{Roman}'s High Latitude Time Domain Survey (HLTDS) and the \textit{Vera C. Rubin} Observatory's Legacy Survey of Space and Time (LSST).}
   {We employed two approaches: one focusing on known multiply imaged galaxies (arcs) behind cluster fields, along with the SN rates specific to those galaxies (arc-specific), while the second is based on the expected number of lensed SNe exploding in a given volume behind a galaxy cluster (volumetric). We collected all the clusters in the literature that feature a) a well-constrained lens model and b) multiply imaged galaxies behind clusters with high-quality data for the multiply imaged galaxies behind clusters. This allowed us to determine the supernova rate for each galaxy. We provide predictions for 46 clusters visible to the \textit{Vera C. Rubin} Observatory, as well as for 9 observable by \textit{Roman}’s HLTDS, depending on whether the clusters fall within the survey’s observing field. }
   {We predict that the number of multiply imaged SNe discovered by LSST in its first three years is $3.95 \pm 0.89$ from the first approach or $4.94\pm1.02$ from the second. 
   Based on the current proposed observing strategy for the HLTDS, which specifies the requirements on galactic and ecliptic latitudes, the expected number of multiply imaged supernovae ranges from $0.38 \pm 0.15$ to $5.2 \pm 2.2$, depending on the specific cluster observed. However, the exact fields to be targeted remain a matter of discussion.}
  {We conclude that LSST offers great prospects for detecting multiply imaged SNe. If adequate follow-up campaigns are conducted, these capabilities will enable measurements of cosmological parameters independent of conventional probes. These predictions are effectively lower limits, as we only considered the most massive and well-studied clusters in the present work. Here, we provide a recommendation for HLTDS observing field selection, namely:  either MACS J0553.4-3342 or Abell 1758a should be observed by the survey to maximize the number of potential multiply imaged SN discoveries.}

   \keywords{Gravitational lensing: strong; cosmology: observations, supernovae: general, galaxies: star formation; techniques: photometric; galaxies: clusters: general
               }
\titlerunning{Cluster-lensed supernovae with LSST and \textit{Roman}}
   \maketitle
%

\section{Introduction}

The idea that supernovae (SNe) can be found in fields that are lensed by galaxy clusters is not new \citep[e.g.,][]{1988ApJ...335L...9K,2000MNRAS.319..549S,2002MNRAS.332...37G,2003ApJ...592...17B,2003A&A...405..859G}. The gravitational lensing effect, a phenomenon predicted by Einstein's theory of General Relativity, causes massive galaxy clusters to act as cosmic magnifying glasses, significantly amplifying the light from background sources, such as quasars, galaxies, or distant supernovae (SNe) occurring in those galaxies. This effect amplifies light from lensed sources and, in special cases, causes multiple images of the source to appear in several positions around the lens, with a delay in the signal observed between images. \cite{1964MNRAS.128..307R} proposed a method for probing the Hubble constant, $H_0$, through observations of multiply imaged SNe. By obtaining a light curve for each image, time delays between images can be determined, providing constraints on $H_0$. Robust measurements of the time delay can also be used to constrain other cosmological parameters, such as the dark energy equation of state \citep[e.g.,][]{2002A&A...393...25G, 2009A&A...507L..49P,2011PhRvD..84l3529L,2016A&ARv..24...11T}. This method of probing cosmological parameters is known as time-delay cosmography \citep[e.g.,][]{2022A&ARv..30....8T}. Such measurements may prove invaluable, given the $>5\sigma$ disagreement between late-Universe measurements of $H_0$ from the SH0ES program \citep{2022ApJ...938...36R} and early-Universe measurements from the Planck satellite \citep{2020A&A...641A...6P}.

Historically, strongly lensed quasars have been used for time-delay cosmography. The latest result from H0LiCOW by \cite{2020MNRAS.498.1420W}, which combined six lensed quasars, achieved a precision of 
2.4\%, consistent with the Type Ia SN-based prediction by \cite{2022ApJ...938...36R}. However, using SNe offers several advantages over quasars. SNe have predictable light curves, vastly simplifying time delay measurements compared to the stochastic nature of quasars. Additionally, SNe fade quickly, enabling precise photometry through background subtraction once the SN fades. This also enables predictive experiments on the timing and brightness of delayed trailing images, as both SN host fluxes and quasar fluxes tend to be highly blended \citep{2021MNRAS.504.5621D}. Time delay measurements for SNe require much shorter observational campaigns and the impact of microlensing is partially mitigated \citep{2018MNRAS.473...80T,2019A&A...621A..55B}, with less pronounced chromatic effects \citep{2018MNRAS.478.5081F,2019A&A...631A.161H}. However, microlensing can still be a significant source of uncertainty for systems with low time delays on the order of one day or less \citep{2017Sci...356..291G,2023ApJ...948..115P}.

Given the rarity of multiply imaged SNe, it was not until 2014 that the first multiply imaged SN was discovered \citep{2015Sci...347.1123K}, 50 years after Refsdal's original publication. The SN was dubbed ``SN Refsdal.'' Since then, a handful of other multiply imaged SNe have been discovered: two lensed by individual galaxies, iPTF16geu and SN Zwicky \citep{2017Sci...356..291G,2023NatAs...7.1098G}, and six by galaxy clusters \citep{2021NatAs...5.1118R,2022Natur.611..256C,2022TNSAN.169....1K,2024ApJ...961..171F,2024ApJ...967L..37P}. From this set, time delays of the multiple images have been measured for only two cluster-lensed SNe, SN Refsdal and SN H0pe \citep{2024ApJ...967...50P,2025ApJ...979...13P}, with similar analysis ongoing for SN Encore \citep{2024ApJ...967L..37P}. 
Observations of SN Refsdal have enabled a measurement of $H_0$ with a $\sim 6\%$ precision in flat
$\Lambda$CDM cosmology \citep{2018ApJ...860...94G} and in an open $w$CDM model \citep{2024A&A...684L..23G}. Through observations of SN H0pe, with a remarkable redshift of $z=1.78$, it was possible to constrain $H_0$ to be $75.4^{+8.1}_{-5.5}{\rm km~s}^{-1}{\rm Mpc}^{-1}$ \citep{2025ApJ...979...13P}. SN Encore is expected to produce a similar $H_0$ uncertainty of $\sim10\%$, which will be presented by Pierel et al. (in preparation). The expected time delays from the galaxy-lensed SNe, iPTF16geu and SN Zwicky, were less than a day; hence, it was not possible to measure $H_0$ \citep{2020MNRAS.491.2639D,2023ApJ...948..115P}. Lensing by galaxy clusters typically results in longer time delays, ranging from months to years, between multiple images \citep[e.g.,][]{2018A&A...614A.103P,2018ARep...62..917P}, compared to galaxy-scale lenses, which have typical time delays on the order of days or weeks  \citep[e.g.,][]{2024MNRAS.531.3509A}. 

Longer time delays are beneficial because they reduce the impact of microlensing on $H_0$ measurements, for which one of the sources of uncertainty is that of the time delay. Typically, microlensing introduces an uncertainty on the order of magnitude of one day, which results in a larger relative error for lower time delays. At the extreme, the microlensing uncertainty can be on the same order of magnitude as the measured time delay, making measuring $H_0$ unviable, as was the case with SN Zwicky \citep{2023ApJ...948..115P}.

It is particularly valuable in the lensing scenario to study SNe Type Ia (SNe Ia), which are used as ``standardizable'' candles to measure cosmological parameters \citep[e.g..][]{1998AJ....116.1009R,1999ApJ...517..565P}. Their standardizable absolute magnitude, coupled with their well-understood light curve evolution \citep[e.g.,][]{2007ApJ...663.1187H,2021ApJ...923..265K,2022ApJ...939...11P}, can provide constraints on lens modelling and break the mass-sheet degeneracy \citep{1985ApJ...289L...1F,2003MNRAS.338L..25O}; that is, if the mililensing and microlensing effects are not extremely strong \citep[e.g.,][]{2020MNRAS.491.2639D,2024ApJ...967...50P}. SN Ia magnifications could also be used as inputs for cluster lens models, providing valuable constraints in regions lacking traditional strong- and weak-lensing information. For example, \cite{Rodney2015} tested 17 gravitational lens models of the galaxy cluster Abell 2744 by comparing the measured magnification of a lensed SN Ia to magnifications predicted by those models, providing valuable input for both  the models of this cluster and to cluster modeling methodology in general.

    Strong lensing also allows for the detection of SNe that would otherwise be too dim to observe without magnification, enabling the study of high-redshift SNe that are otherwise inaccessible. For example, \cite{2016A&A...594A..54P} sets limits on the volumetric rates of the core collapse SN (CC SN) up to $z \approx 2.5$ where they are poorly understood. Since CC SNe are tracers of star formation, studying their rates allows for the cosmic star formation history to be probed \citep[cSFH, see e.g.,][]{2015ApJ...813...93S,2020ApJ...890..140S}. Thus, strong lensing of SNe can help probe cSFH by enabling the detection of high-redshift CC SNe. Lastly, strong lensing enables studying spectra of high-redshift SNe, and testing for redshift evolution, namely, the dependence of SN characteristics, such as spectra or color evolution, on redshift \citep[e.g.,][]{2017A&A...603A.136P,2024ApJ...970..102C,2024ApJ...967L..37P}. Thus, analyzing a single strongly magnified high-redshift SN offers valuable insights for both astrophysical and cosmological applications. 

The \textit{Vera C. Rubin} Observatory is scheduled to have its first light in 2025 and to commence its Legacy Survey of Space and Time \citep[LSST;][]{2019ApJ...873..111I}. The survey will scan the sky at an unprecedented rate at high depth in six filters (\emph{ugrizy}), ranging from $22.7~\rm{mag}$-$24.5~\rm{mag}$, depending on the filter, discovering numerous transients. The \textit{Nancy Grace Roman} Space Telescope \citep[\textit{Roman};][]{2019arXiv190205569A} is scheduled to launch in the second half of the decade. This $2.4$~m space telescope has a large $0.281$~square degree field of view and is capable of near-infrared imaging and slitless spectroscopy. The telescope is going to conduct numerous surveys for different science cases. Of particular interest for strongly lensed SNe is the High Latitude Time Domain Survey (HLTDS). The observing strategy of the \textit{Roman} is currently under discussion and \cite{2021arXiv211103081R} provided a proposal for the HLTDS. These include 1) a high ecliptic latitude ($|\beta| > 54^\circ$) to ensure continuous visibility throughout the year, and 2) a high galactic latitude $|b|$ to minimize the effects of galactic extinction. The strategy also proposes a five-day cadence and very deep observations, achieving depths between $25.4$~mag-$26.7$~mag, depending on the survey tier and filter. As the exact observing fields for the survey have not yet been selected, a detailed analysis of \textit{Roman}'s strongly lensed SN discovery potential may be used in order to form recommendations for observing field selection, taking into account the HLTDS science and technical specifications. 

The estimates for galaxy-lensed SNe in LSST were previously studied \citep[e.g.,][]{2010MNRAS.405.2579O,2013MNRAS.429.2392K,2019ApJS..243....6G, 2019A&A...631A.161H,2019MNRAS.487.3342W,2024MNRAS.531.3509A}, predicting hundreds of lensed SNe over the 10 years of the LSST duration. However, only a small portion of those will be useful for achieving competitive $1.3\%$ precision in $H_0$ measurements in the flat $\Lambda$CDM cosmology, assuming that follow-up observations are conducted \citep{2020A&A...644A.162S}. For \textit{Roman}, \cite{2021ApJ...908..190P} predicts the discovery of $\sim11$ lensed SNe Ia and $\sim20$ CC SNe, depending on the survey strategy, over the two years of the HLTDS. For cluster-lensed SNe, the expectations for LSST have already been presented in \cite{2020Symm...12.1966P}. However, this study only considered the spectroscopically confirmed and multiply imaged galaxies within the fields of view of five massive lens galaxy clusters, which comprises only a small fraction of clusters that LSST is poised to observe.

In this paper, we explore the discovery potential for cluster-scale lensed SNe of the two upcoming surveys,  LSST and HLTDS, using a large, updated sample of well-studied clusters. We employ two approaches: one focusing on known multiply imaged galaxies (arcs) and the SN rates specific to those galaxies (arc-specific) and another based on the expected number of lensed SNe exploding in a given volume behind a galaxy cluster (volumetric). 

The paper is structured as follows. First, in Sect.~\ref{sec:sample}, we present the cluster sample studied in this paper. Then in Sect.~\ref{sec:methodology}, we elaborate on the methods we used to obtain the expected number of lensed SNe in the two surveys. The results are shown in Sect.~\ref{sec:results}. We discuss our results in Sect.~\ref{sec:discussion} and present our results in Sect.~\ref{sec:conclusion}.

\section{Galaxy cluster sample} \label{sec:sample}
To obtain the SN yields in the aforementioned surveys, we constructed two datasets for the different estimation methods we employed. These datasets are described below. 

\begin{figure*}[h]
\centering
    \centering
    \begin{subfigure}[b]{0.9\textwidth}
    \includegraphics[width=\textwidth]{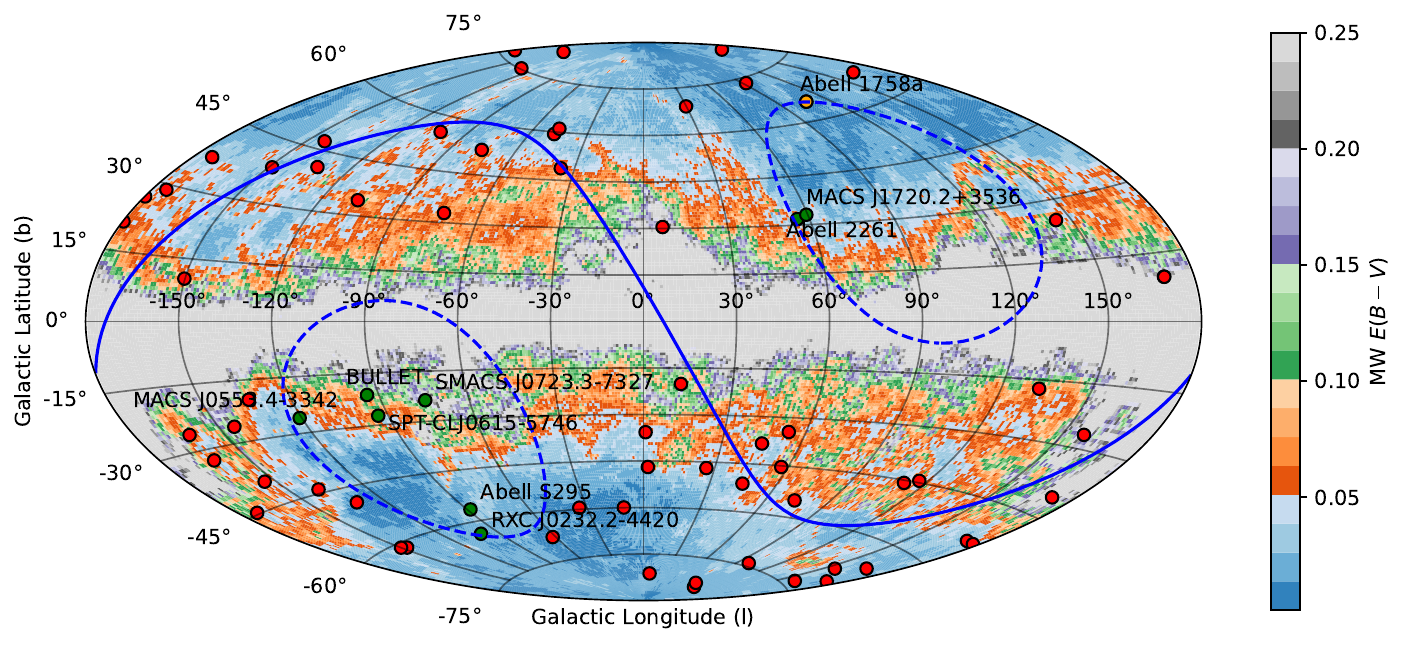}
    \end{subfigure}

\caption{Clusters in our sample overlaid with considerations for the HLTDS observing fields, as proposed by \cite{2021arXiv211103081R}. Red points indicate clusters for which continuous viewing is not possible due to their ecliptic latitude $|\beta| < 54^\circ$. Green points indicate clusters which fulfill this condition, marked by the blue dashed line. The continuous blue line indicates the ecliptic plane. The cluster Abell 1758a is marked orange as a special case, requiring only a minor relaxation of the $\beta$ constraint (see Sect. \ref{sec:res_volum}). The Milky Way extinction map is from \cite{1998ApJ...500..525S}.}
\label{fig:extinction_roman}
\end{figure*}

\subsection{Data sample for the arc-specific method}

In the arc-specific estimation method, the aim was to estimate the SN yields in the known arcs behind galaxy clusters.  This requires determining SN rates within those arcs, which can be derived from their star formation rates (SFRs) and stellar masses \citep{2011MNRAS.412.1473L}. These properties, in turn, can be inferred from the spectral energy distributions (SEDs) of the arcs. Estimating SFRs from SED fittings can be very challenging because of the age-dust-metallicity degeneracy, 
unless high-quality data are available \citep[e.g.,][]{2013ARA&A..51..393C}. Therefore, we required that the arcs have photometry in a wide wavelength range that can be provided from surveys conducted with \textit{Hubble} Space Telescope (HST), \textit{James Webb} Space Telescope \citep[JWST;][]{2024ApJ...974...92B}, \textit{Spitzer} Space Telescope (\textit{Spitzer}) photometry \cite{2022ApJS..263...38K}, or ALMA Lensing Cluster Survey\footnote{\url{http://www.ioa.s.u-tokyo.ac.jp/ALCS/}, date of access: 05.12.2023} \citep[ALCS;][]{2024ApJS..275...36F}. We searched for published lists of arcs behind clusters that have multi-wavelength photometric catalogues and well-constrained magnification from high quality lens models with many spectroscopically confirmed multiply imaged galaxies. There are 19 clusters satisfy these criteria, comprising a total of 872 arcs. However, observability constraints due to the position of the clusters in the sky reduce this sample to 16 clusters visible to LSST, with only one observable by HLTDS according to the proposed strategy by \cite{2021arXiv211103081R}. These clusters have remarkable photometric and spectroscopic datasets taken from large programs, such as   JWST  Ultradeep NIRSpec and NIRCam ObserVations before the Epoch of Reionization \citep[UNCOVER;][]{2024ApJ...974...92B,2024ApJ...976..101S}, Reionization Lensing Cluster Survey \citep[RELICS;][]{2018ApJ...859..159C,2020ApJ...889..189S},  \textit{Hubble} Frontier Fields \citep[HFF;][]{2017ApJ...837...97L}, Cluster Lensing And Supernova survey with \textit{Hubble} \citep[CLASH;][]{2012ApJS..199...25P} programs, and Multi Unit Spectroscopic Explorer (MUSE) observations \cite{2021AnA...646A..83R}. The list of clusters which were analyzed in this manner is shown in Table \ref{tab:cl_list}, column ``Phot. sample.'' We note that 67\% of arcs in our sample have spectroscopically confirmed redshift. For arcs that do not have spectroscopic redshift, we used estimates obtained from the lens modeling process. Those values are derived from high quality lens models, based on a photometric redshift prior from broadband photometry and are therefore well-constrained for recently published, high-quality cluster lens models. 

\subsection{Data sample for the volumetric method}

In the volumetric estimate method, we used the gravitational lensing model of a galaxy cluster to find the volume behind the cluster in which sources can form multiple images. Then, we combined it with a volumetric SN rate to estimate the expected number of SNe behind a cluster and to compute SN yields. Therefore, we surveyed the literature for galaxy clusters with well-constrained, publicly available lensing models. The search yielded models of 71 galaxy clusters, listed in Table \ref{tab:cl_list}. These clusters formed our sample for the volumetric SN yield estimate method. In cases where multiple models were publicly available, we selected the most recently published or uploaded model, as we expect it to be made with the most recent and comprehensive data. Out of those, 46 are visible to LSST, and 9 fulfill or are close to fulfilling the HLTDS requirement of ecliptic latitude $|\beta|>54^\circ$ proposed by \cite{2021arXiv211103081R}. Figure \ref{fig:extinction_roman} shows the distribution of clusters in our sample and their observability for the \textit{Roman}'s HLTDS due to ecliptic latitude and Milky Way extinction constraints.

\section{Methodology}
\label{sec:methodology}

In this section, we describe the methodology and assumptions used in our two different approaches.
We first present the method of estimating the SN yields in the known arcs from their specific SN rates, which are based on the specific SFRs and total stellar masses formed in those galaxies $M_\star$. Then we describe the volumetric method, which is based on the lensed volume behind clusters and an assumed volumetric SN rate.

Throughout this work, we assume a flat $\Lambda$CDM Universe model with $H_0 = 70~\rm km~s^{-1}Mpc^{-1}$, and matter and dark energy density fractions of  $\Omega_{\rm m} = 0.3$ and $\Omega_{\Lambda} = 0.7$, respectively.

\subsection{SN yields in the known arcs}

When observing individual galaxies, the expected number of detected SNe of subtype $j$, $N_j$,  depends on the survey control time for a given galaxy $T_j(z, \mu, ext)$, as well as the specific SN rate $R_j^{\rm s}$ in that galaxy,
\begin{equation}
\label{eq:no_of_sne}
    N_j = T_j(z, \mu, ext) \frac{R_j^{\rm s}}{1+z},
\end{equation}
with the factor $(1+z)$ converting the rest-frame SN rate to an observer-frame rate. The survey control time indicates the total amount of time that a survey is sensitive to detecting an SN at a given redshift $z$, affected by extinction \textit{ext}, and can be understood as the equivalent time of continuous observation sufficiently deep to detect an SN. We note that due to lensing, $T_j(z, \mu, ext)$ depends on magnification and it will therefore  be different for individual images of the same galaxy. The survey control time is further discussed in detail in Sect. \ref{sec:control_time}. For every multiply imaged galaxy modeled in Sect. \ref{sec:cigale}, we computed the specific SN rates, as outlined in Sect. \ref{sec:sn_rates}. When the expected SN yield in an image were calculated, we considered each image of a galaxy separately and independently of other images and made no assumptions for time delays between images. For a given galaxy system, we took an average over all images in summary statistics. 

\subsubsection{Specific SN rates in the known arcs}
\label{sec:sn_rates}

The progenitors of CC SNe are massive stars and, as a result, their short lifespans are negligible compared to the timescales that govern the change in SFRs specific to a given galaxy. In other words,  the specific CC SNe would be expected to trace the locally recent SFR. Here, we have assumed that the stellar initial mass function (IMF) is constant throughout the Universe and that the mass range for SN progenitors is constant. Therefore, the fraction of stars $k_{\rm CC}$ which explode as SNe, is a universal constant in our analysis. We used the value $k_{\rm CC} = 0.0091 \pm 0.0017M_\odot^{-1}$ as determined from observations by \cite{2015ApJ...813...93S}. Thus, the specific CC SN rate in a given galaxy, measured in ${\rm yr}^{-1}$, is
\begin{equation}
    R_{\rm CC}^{\rm s} = k_{\rm CC} \cdot \rm SFR.
\end{equation}

We note that the $k_{\rm CC}$ provided by \cite{2015ApJ...813...93S} differs from a value computed from a \cite{1955ApJ...121..161S} IMF, as it is a fit to observational data.  A $k_{\rm CC}$ derived from a \cite{1955ApJ...121..161S} IMF would be about $25\%$ lower but within a reasonable range of uncertainty \citep{2012ApJ...757...70D,2015ApJ...813...93S}.

We further divide CC SNe into seven subtypes: IIP, IIL, IIn, IIb, Ib, Ic, and Ic-BL, following \cite{2019MNRAS.489.5802V}, who combined relative SN rates from \cite{2017PASP..129e4201S} with Type IIL and Type IIP relative rates from \cite{2011MNRAS.412.1441L}. These relative rates are volume-limited in their respective samples. Thus, for a CC SN subtype $j$, the specific SN rate is
\begin{equation}
        R_j^{\rm s} = k_{\rm CC} \cdot \mathrm{SFR} \cdot r_j,
\end{equation}
where $r_j$ is the relative rate of the subtype $j$ of CC SNe, such that $\Sigma r_j = 1$. We assumed that these relative rates are constant through cosmic history.

The relation between specific SN Ia rates and galaxy properties is more complex. The two main models of their progenitors necessitate the formation of at least one white dwarf (WD), which is a slower process than the lifespan of CC SN progenitors due to lower WD progenitors' masses. After the formation of the WD, additional time is required for the infall of matter onto it,  either coming from a companion star or taking place during an inspiral period in a binary system. To calculate the specific Type Ia SN rate $R^{\rm s}_{\rm Ia}$, we used the simple model from \cite{2005ApJ...629L..85S}, which provides a practical choice for our purposes. According to this model, $R^{\rm s}_{\rm Ia}$ consists of two components: a prompt element proportional to SFR, and a time-extended element proportional to the total stellar mass formed within the galaxy:
\begin{equation}
    R^{\rm s}_{\rm Ia} = A \cdot M_\star + B \cdot \mathrm{SFR},
\end{equation}
with $A$ and $B$ being the proportionality constants. We used updated values from \cite{2018MNRAS.480...68A}:
\begin{align*}
    A &= (4.66 \pm 0.56) \cdot 10^{-14} ~ \mathrm{SNe~yr}^{-1}M_\odot^{-1}, \\
    B &= (4.88^{+0.54}_{-0.52}) ~ \cdot 10^{-4} \frac{\mathrm{SNe~yr}^{-1}}{M_\odot \mathrm{yr}^{-1}}.
\end{align*}

\subsubsection{Star formation rates and stellar masses of the arcs}
\label{sec:cigale}

To derive the physical properties of the galaxies required for the arc-specific method, we used the latest release of the Code Investigating GALaxy Emission (CIGALE; \citealt{cigale19}). CIGALE is a state-of-the-art SED modelling and fitting code that combines observations from the far-UV to the far-IR and radio. In our sample, we typically have between seven and eight bands of photometry available from HST and \textit{Spitzer} surveys, while for Abell 2744, we added the publicly available JWST data \citep{weaver24,2024ApJ...976..101S}, extending the photometric data to 14 bands in optical to near-infrared regime, up to 8 $\mu$m. For a small subset of objects, we added the available Band 6 (1 mm) ALMA photometry \citep{2024ApJS..275...36F}. We fit the SEDs to the photometry data of all multiply imaged galaxies from our sample with the following procedure. We first demagnified the data using magnification estimates provided (when available) from photometric catalogs in the literature or from cluster models from our sample otherwise. We limited the magnification estimates to a maximum of 100, and omitted magnification uncertainty estimates (as discussed in Sect. \ref{sec:error_budget}).

CIGALE is designed for estimating a wide range of physical parameters by comparing modelled galaxy SEDs to observed ones. For each parameter, CIGALE makes a probability distribution function (PDF) analysis, providing the likelihood-weighted mean of the PDF as the output value, with the associated error being the likelihood-weighted standard deviation. As CIGALE entirely conserves the energy between dust absorption in the UV-to-near-IR domain and emission in the mid-IR and far-IR, we included far-infrared and submillimeter (IR/submm) constraints from ALMA observations whenever possible, to ensure a proper estimate of the SFR and $M_\star$ in sources with significant dust emission \citep[e.g.,][]{donevski20,haskell22}. We carefully chose the model parameters following some of the most recent prescriptions optimized for a wide range of star-forming galaxies over large redshift range; in particular, those observed with HST, JWST, \textit{Spitzer}, and ALMA \citep[e.g.,][]{zou22,ciesla22,wang24}.

To construct the SED model for each individual galaxy, we applied the stellar population synthesis module, nebular module, star-formation history module, and dust attenuation module. To construct the stellar component, we used a \citet{bruzual03} stellar population synthesis model, together with a \cite{chabrier03} IMF. We use the grid of metallicities and limited the maximal value to solar, in line with observations (e.g., \citealt{sanders19}; \citealt{sanders24}). We adopted the flexible star-formation history (SFH), which is composed of a delayed component with an additional episode of a star formation burst. The functional form is given as

\begin{equation}
\rm SFR(\mathit{t}) = \rm SFR_{\rm delayed}(\mathit{t})+\rm SFR_{\rm burst}(\mathit{t}),
\end{equation}
where $\rm SFR_{\rm delayed}(\mathit{t})\propto \mathit{t} e^{-t/\tau_{\rm main}}$, and $\rm SFR_{\rm burst}(\mathit{t}) \propto e^{-(t-t_{0})/\tau_{\rm burst}}$. Here, $\tau_{\rm main}$ represents the e-folding time of the main stellar population, while $\tau_{\rm burst}$ represents e-folding time of the late starburst. For the main stellar population age we assumed a dense grid of 20 linearly sampled values ranging from 0.8 Gyr to 13 Gyr. Following the prescription from \citep{villa20}, the time for recent burst of constant star formation has been fixed to
70 Myr. The remaining SFH parameter (mass fraction of the late burst, defined as a relative ratio to a total stellar mass built in the recent starburst) is chosen similarly to that of \cite{donevski20} and \cite{ciesla22}, in the context of  galaxies studied over wide range of redshifts. In particular, we allow this parameter to be 0, 0.01, 0.1, and 0.2. In particular, our choice of SFH is motivated by the study by \cite{ciesla17} (see also \citealt{forrest18}), who demonstrated that a flexible delayed+burst SFH model is able to consistently model the observations with respect to the $\mathrm{SFR}-M_{\star}$ plane. We also add the module for nebular emission, which uses the template from \citep{villa20}. Following \citep{nanni20}, we keep the ionization parameter fixed ($U =-1.5$), while probing two gas-metallicities (solar and $4\times$ below solar).

To model the dust attenuation we adopt a double power-law recipe described in \cite{cf00}. The \cite{cf00} attenuation law assumes that birth clouds (BCs) and the interstellar medium (ISM) each attenuate light according to fixed power-law attenuation curves. The formalism is based on age-dependent attenuation, meaning that a differential attenuation between young (age $<10^7\:\rm yr$) and old (age $>10^7\:\rm yr$) stars is assumed. Both attenuation laws are modelled by a power-law function, with the amount of attenuation quantified by the attenuation in the V band. We chose to probe the combination of widely used values for power-law slopes (BC and ISM), namely, -0.44 and -0.7 (e.g., \citealt{ciesla21}; \citealt{hamed23}). We allowed for other input parameters (V-band attenuation in the interstellar medium (ISM) and attenuation in the birth clouds) to vary, as suggested in \cite{donevski20} and \cite{ciesla22}.

The procedure outlined above resulted in 4,280,000 models in the parameter space grid. The number of models in CIGALE reflects the number of generated SEDs multiplied by the number of galaxies.
The grid of models (fluxes and physical properties) is estimated over all the possible combinations as an input (SFH, stellar population synthesis model, attenuation and nebular modules).  In these modules, we fit seven parameters in total (e-folding time of the main stellar population model, age of the main stellar population in the galaxy, starburst  mass fraction, metallicity, and dust attenuation in the ISM and the birth clouds). The remaining parameters were fixed, as detailed above. We confirmed that the fit SEDs are of a good quality, which is quantified with a median $\chi^{2}/{\rm DoF} = 1.9$, with only a few outliers with  $\chi^{2}/{\rm DoF} > 10$, which we rejected. 

 In addition, we followed the approach introduced in recent CIGALE studies \citep[e.g.,][]{ciesla22, donevski23} and performed a ``mock'' test to analyze the ability of the code to constrain the key parameters of this study. To achieve this goal, we used the functionality available in CIGALE to create a mock catalogue of objects for each galaxy for which the physical parameters are known. The best SED model of each catalogue galaxy was then integrated into the same sets of filters as our observed sample, and the fluxes were perturbed by adding noise from a Gaussian distribution which matched the observed catalogue's flux density uncertainty distribution. We recovered a great match between the input and output values. In particular, we found that for SFR and $M_{\star}$ analyzed in this work, the dispersion of recovered values (i.e., the difference on a log-scale between the input physical properties and the best output parameters) follows a normal distribution, with $\sim82\%$ of sources lying within the mean offset from zero of only $\pm0.2\:\rm dex$. This test allows us to conclude that available data provide a reliable constraint on SFR and $M_{\star}$ for a vast majority of our galaxies. The typical resulting uncertainties for SFR and $M_\star$ are $14\%$ and $8\%$, respectively,  remaining consistent across redshifts.

 In our sample, individual images of the same system can have different multiwavelength coverage; for instance, when an image is less magnified and too faint to be detected or it is close to a foreground source. This leads to situations where some images have insufficient data to constrain the lensed galaxy's physical parameters. Therefore, we modeled the SED for each image of the system independently and then we chose the image with the lowest total uncertainty on SFR and $M_\star$ to be representative of the entire arc system. 
 
Of the 872 arcs in our sample, 749 arcs belonging to 296 systems had an available IR photometric constraint from ALMA or \textit{Spitzer}, fulfilling our condition to attempt the modeling process. However, for most arcs, the available photometry was insufficient to constrain the physical properties of the galaxy, which was signified by the fit not converging to a specific value, and uncertainties on $\mathrm{SFR}$ and $M_\star$ being arbitrarily high. We find that a cut on 50\% uncertainty on $\mathrm{SFR}$ and $M_\star$ allowed us to separate the arcs for which no constraint could be found from the well-constrained sample.

After this cut, 137 arcs with well constrained SFHs remained. However, we rejected three images for which the results were unlikely high, namely, $\mathrm{SFR} > 10^3 M_\odot \mathrm{yr}^{-1}$, leaving 134 well constrained arcs. By extrapolating the SFHs to other arcs belonging to the same systems, we obtained 285 arcs which have a well constrained SFH, belonging to 90 systems. Of those, 143 arcs are at redshift $z < 3$ in the 16 clusters within LSST's observing fields, and 12 arcs are within the proposed observing fields of HLTDS.

\subsubsection{Survey control time}
\label{sec:control_time}

\begin{figure*}[t]
\centering
    \centering
    \begin{subfigure}[b]{\textwidth}
    \includegraphics[width=\textwidth]{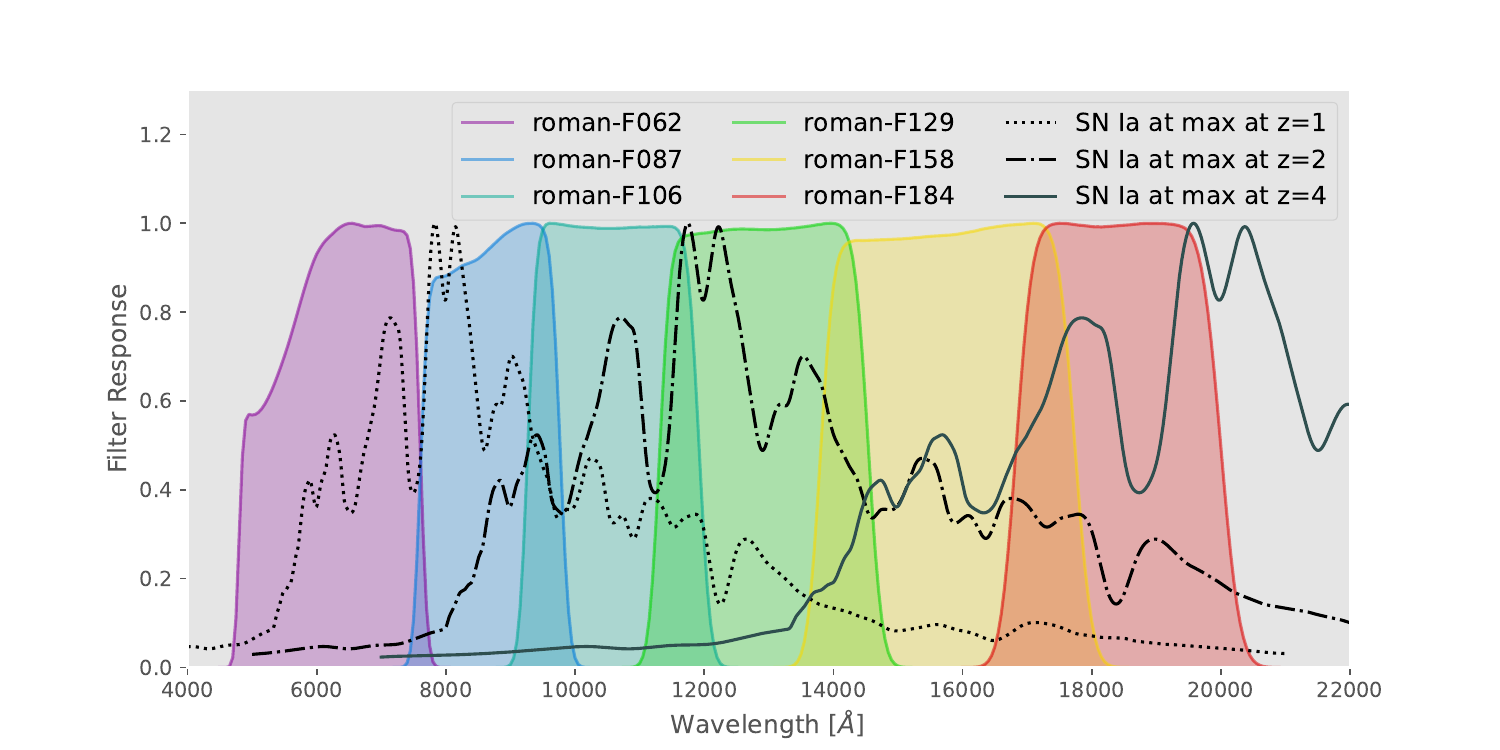}
    \end{subfigure}

\caption{Response curves for \textit{Roman}'s HLTDS overlaid with typical spectra of SNe Ia at three different redshifts:  $z = 1$, $z = 2$, and $z = 4$. The Wide tier of HLTDS will use four filters: F062/R, F087/Z, F106/Y, and F129/J, while the Deep tier will use the filters F106/Y, F129/J, F158/H, and F184/F. From the figure it is evident that HLTDS, particularly the Deep tier, is well-suited for discovering high-redshift SN Ia, even beyond $z = 4$.}
\label{fig:roman_filters}
\end{figure*}

\begin{figure*}[t]
\centering
    \centering
    \begin{subfigure}[b]{\textwidth}
    \includegraphics[width=\textwidth]{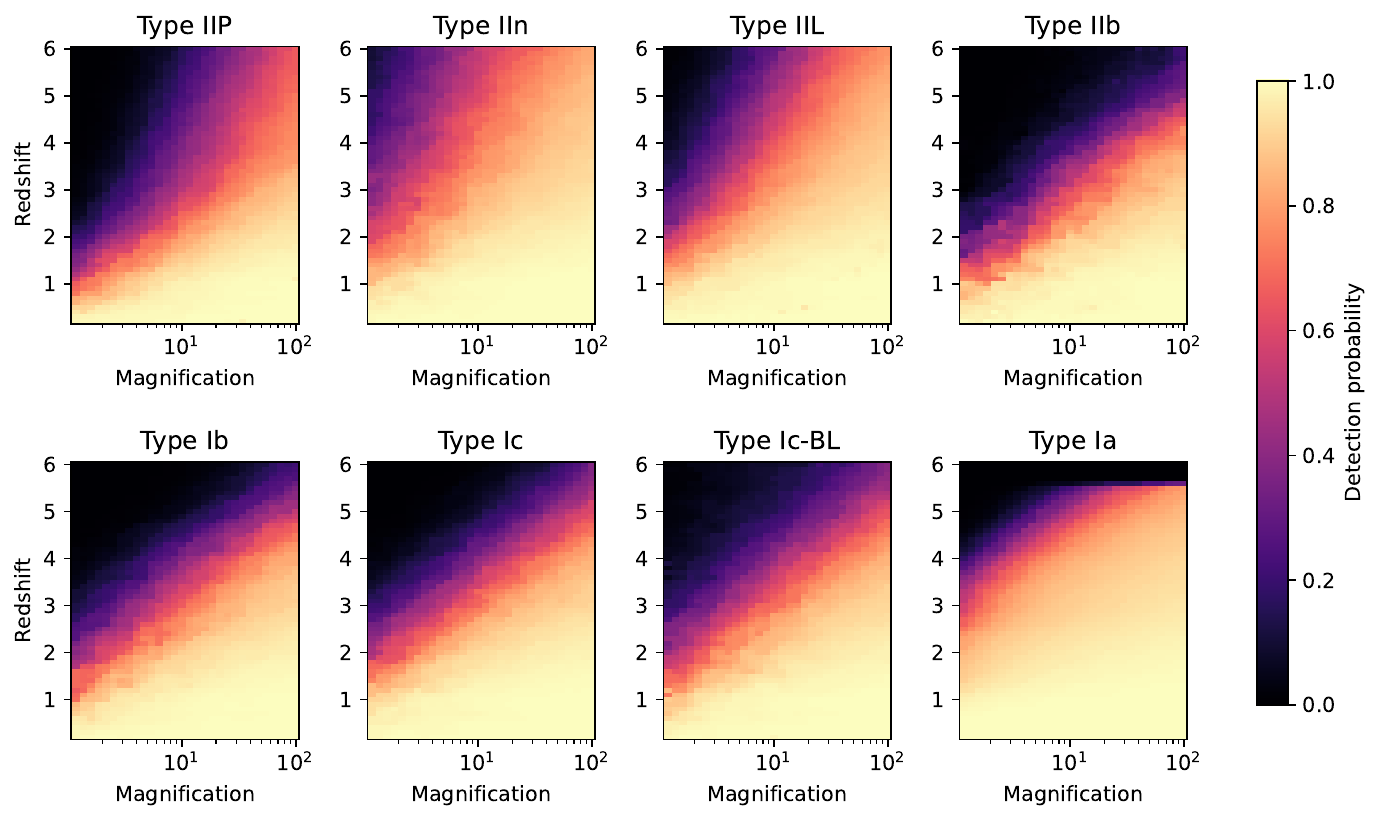}
    \end{subfigure}

        \caption{Detection probability of an SN of a given subtype as a function of magnification $\mu$ and redshift $z$ in the Deep survey tier of HLTDS, using sampled host extinction parameters.}
\label{fig:detrates_roman_deep}

\end{figure*}

\begin{figure*}[t]

    \centering
    \begin{subfigure}[b]{\textwidth}
    \includegraphics[width=\textwidth]{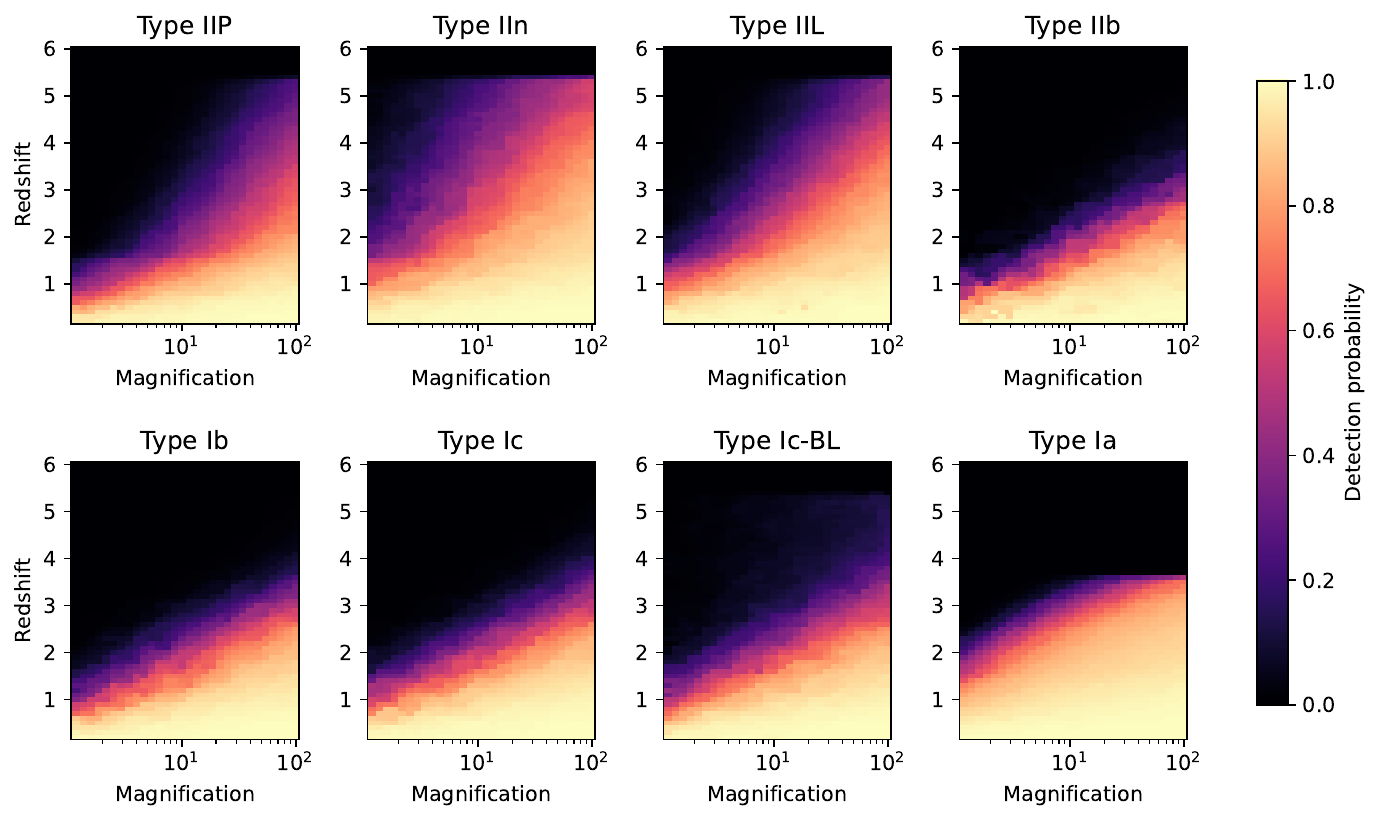}
    \end{subfigure}
    
    \caption{Detection probability of an SN of a given subtype as a function of magnification $\mu$ and redshift $z$ in the Wide survey tier of HLTDS, using sampled host extinction parameters.}
\label{fig:detrates_roman_wide}
\end{figure*}

\begin{figure*}[t]
\centering
    \centering
    \begin{subfigure}[b]{\textwidth}
    \includegraphics[width=\textwidth]{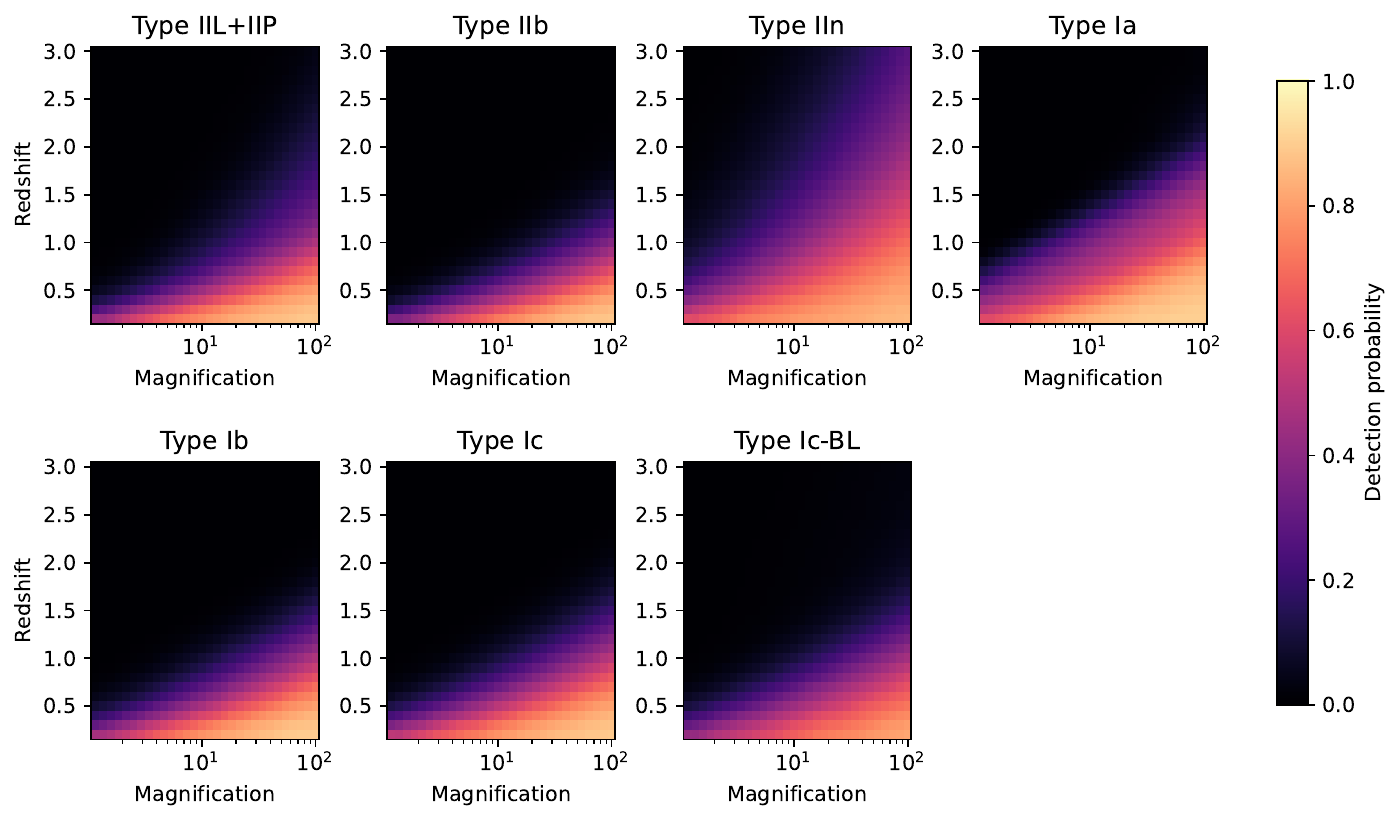}
    \end{subfigure}

    \caption{Detection probability of an SN of a given subtype as a function of magnification, $\mu,$ and redshift, $z,$ by LSST, in the baseline survey strategy, using sampled host extinction parameters. The probabilities shown are averaged over all analyzed clusters. Note that the detection probability takes into account SNe which explode while a given field is not observable due to seasonal constraints.}
\label{fig:detrates_lsst}
\end{figure*}

Survey control time, $T_j(z, \mu, ext)$, is defined as the time during which a survey is sensitive to an SN. It can be expressed as the product of the total observation time, and the probability of detecting an SN that explodes in that time-frame. To account for SNe that explode before the start of the survey or peak after its completion, but are still detectable, we would need to simulate a longer time period $t_{\rm sim}$ than the survey duration. Therefore, the survey control time is 

\begin{equation}
\label{eq:surver_control_time}
   T_j(z, \mu, ext) = t_{\rm sim} \cdot p_j(z, \mu, ext),
\end{equation}

\noindent where $t_{\rm sim}$ is the total time for which we simulate SN light curves, and $p_j(z, \mu, ext)$ is the probability that an SN of type $j$, at redshift, $z$, magnified by a factor of $\mu$, will be detected, if it explodes in the simulated period, taking into account extinction \textit{ext} of both the host galaxy and the Milky Way. Note that the detection probability depends on the SN luminosity function of the specific SN type, as will be discussed in Sect. \ref{sec:snana_hltds}. We interpret the fraction of SNe detected in our simulations to be equal to $p_j(z, \mu, ext)$.

We used the Supernova Analysis Package \citep[SNANA;][]{2009PASP..121.1028K} to simulate the SN light curves, as observed by telescopic surveys, and calculate their detection probabilities. SNANA is the most commonly used code for simulating SN lightcurves due to its speed, accuracy and flexibility \citep{2021ApJ...908..190P}. 

\subsubsection{SNANA simulations for HLTDS}
\label{sec:snana_hltds}

The proposed survey strategy for the two-year \textit{Roman} HLTDS, as outlined by \cite{2021arXiv211103081R}, employs a dual-tier observational design to balance broad sky coverage with greater depth in targeted areas. This approach consists of a Wide tier, covering approximately 19.04 square degrees using four filters, F062/R, F087/Z, F106/Y, and F129/J, which span wavelengths from 0.48 to 1.454 $\mu$m. The Deep tier focuses on a narrower area of about 4.2 square degrees, utilizing the filters F106/Y, F129/J, F158/H, and F184/F, covering wavelengths from 0.927 to 2.00 $\mu$m. A key feature of this strategy is the five-day cadence maintained throughout the two-year observational period. We note that \textit{Roman} will also have slitless spectroscopy, but here we only consider the photometric part of the survey. 

In our simulations, we considered both the Wide and the Deep survey tier. Following \cite{2021arXiv211103081R}, for the Wide tier, we simulated observations in the filters F062/R, F087/Z, F106/Y, and F129/J with $100$~s exposure times, resulting in limiting magnitudes of $26.4$~mag, $25.6$~mag, $25.5$~mag, and $25.4$~mag respectively. For the Deep tier, we simulated observations in F106/Y, F129/J, F158/H, and F184/F at $300$~s exposure times, resulting in limiting magnitudes of $26.7$~mag, $26.6$~mag, $26.5$~mag, and $26.7$~mag, respectively. The observations were spaced equally with a constant cadence of 5 days over a survey duration of two years. The response curves of filters proposed for HLTDS, overlaid with example SN Ia spectra at redshifts $z = 1$, $z=2$, and $z=4$, are shown in Fig. \ref{fig:roman_filters}. 

As the focus of this paper is on SNe that are magnified, we simulated SNe on a grid, in 30 bins spaced equally on a logarithmic scale, with a maximum magnification value of 100,   resulting in a range of $\mu \in [1.166, 100]$, and  at redshifts in the range of $z \in [0.2, 6]$ in bins of $0.1$, of the following types:  Ia, Ib, Ic, IIb, IIn, IIL, IIP, and Ic-BL. For each SN type, we simulated at least 200 SNe for every possible magnification and redshift in the grid. We simulated a survey of 725 days of observations in accordance with the survey specifications proposed by \cite{2021arXiv211103081R}, and simulated SNe with peaks which occur between 50 days before the first observation, to 15 days after the last observation of a given field. Thus, the total time in which SNe are simulated is $t_{\rm sim} = 790$~days. We chose the detection threshold to be two observations in any filters above a $5\sigma$ signal-to-noise ratio (S/N).

For SN Ia simulations, we used the SALT3 SED model \citep{2021ApJ...923..265K} extended to near-infrared by \cite{2022ApJ...939...11P}, with asymmetric Gaussian parameters from \cite{2016ApJ...822L..35S}, specfically using Table 1, row ``G10 high-$z$'' \citep{2010A&A...523A...7G}.  As host extinction is included in the SED model, we did not simulate additional host extinction in SNANA.

In CC SN simulations, we used spectrophotometric templates presented by \cite{2019MNRAS.489.5802V}, and we adjusted the SEDs to match luminosity functions given by \cite{2014AJ....147..118R}. Both the spectrophotometric templates and luminosity functions were provided in a de-reddened state by the respective authors, meaning they included corrections for extinction from both the Milky Way and the host galaxy.
To simulate the observed light curves, we applied host galaxy extinction to the de-reddened SEDs using a Milky Way-like extinction law \citep{1999PASP..111...63F}. We applied this to each light curve, and verified whether the reddened light curve is still beyond the ${\rm S/N} > 5$ detection threshold. We used a fixed value of $R_V = 3.1$, while the $A_V$ value is assumed to be the best fit value of the CIGALE SED modeling (Sect 3.1.2).

For the volumetric estimate (see Sect. \ref{sec:volum_estimate}), where individual host extinctions are not as readily available, we sampled host extinction parameters from the prior probability distribution for the SN host galaxy extinction for CC SNe provided by \cite{2014AJ....148...13R}, Fig. 7, ``mid'' distribution, which corresponds to an $R_V = 3.1$ and an $A_V$ sampled randomly from the sum of a Gaussian and exponential distribution over $[0, \infty]$. The Gaussian component's dispersion is $\sigma = 0.6$, and the exponential component is of the form $e^{-A_V/\tau}$, where $\tau = 1.7$. The two components were normalized so that the Gaussian component's value at $A_V = 0$ is 4 times that of the exponential component. For each SN which was marked as detected in the SNANA simulations, we sampled and applied host extinction 10 times, and noted the fraction of light curves that remained above the detection threshold.

As the HLTDS survey specification proposed by \cite{2021arXiv211103081R} recommends high galactic latitude targets to minimize Milky Way extinction, and \textit{Roman}'s filter set covers infrared wavelengths, in which extinction is considerably weaker than in the visible spectrum, Milky Way extinction is negligible. To mitigate numerical artifacts, we applied a simple median filter with a 3x3 kernel over each $p(z, \mu)$ plane. Similarly, this was also done for the LSST simulations, which are presented in the next section. The detection probability planes are shown in Figs. \ref{fig:detrates_roman_deep} and \ref{fig:detrates_roman_wide} for the Deep and Wide survey tiers, respectively. 
 
\subsubsection{SNANA simulations for LSST}
\label{sec:snana_lsst}

To obtain the survey control time $T_j(z, \mu, ext)$ for LSST, we used the \textit{SNANA} pipeline created for the ELAsTiCC data challenge\footnote{\url{https://portal.nersc.gov/cfs/lsst/DESC_TD_PUBLIC/ELASTICC/}, date of access: 07.12.2023}. 
Following this pipeline, for CC SN, we used the same luminosity functions as in HLTDS simulations described in Sect. \ref{sec:snana_hltds}, with the exception that we simulate Type IIP and Type IIL SNe in one simulation as ``Type II'', with proportions between the two types being 7.215 Type IIP to 1 Type IIL SN, from \cite{2011MNRAS.412.1441L}. For SN Ia, we used the SALT2 model \citep{2007A&A...466...11G,2010A&A...523A...7G} extended by \cite{2018ApJ...867...23H}, with Asymmetric Gaussian parameters from \cite{2016ApJ...822L..35S}, Table 1, row ``G10 High-$z$'' \citep{2010A&A...523A...7G}. We acknowledge that this is an older model than the new and improved SALT3, however, the photometric difference between the two in the \textit{Vera C. Rubin}'s observatory filter set is expected to be negligible. Thus, we chose to use this model to stay consistent with previous research.

We applied Milky Way extinction to simulated light curves, depending on the lensing cluster's location in the sky. We followed a \cite{1998ApJ...500..525S} dust distribution map updated by \cite{2011ApJ...737..103S}, with a \cite{1999PASP..111...63F} color law. After the simulation, we applied the same host extinction procedure as described in Sect. \ref{sec:snana_hltds}.

We simulated the same period of observations of 1105~days as ELAsTiCC, with cadence corresponding to the \textit{baseline v2.0 10 years} survey strategy\footnote{Opsim run data: \url{http://astro-lsst-01.astro.washington.edu:8081/}; Survey strategy overview: \url{https://www.lsst.org/scientists/survey-design}. Date of access 07.12.2023.} to make use of the existing \textit{SNANA} pipeline for the data challenge. We allowed for SNe with peaks occurring up to 50 days before observations start or 75 days after they end. This resulted in a total $t_{\rm sim}=1230$~days, or $3.37$~years. Note that to calculate a yearly rate, simply dividing the expected values of multiply imaged SNe by 3.37 is a good approximation, but not entirely accurate. For every cluster visible to LSST, we used the filters and observation times assumed by ELAsTiCC at the corresponding area in the sky, and used the same trigger condition of one $5\sigma$ detection in any filter. We simulated 22 million light curves across 46 clusters in LSST's observing field, listed in Table \ref{tab:cl_list}. Similarly to HLTDS simulation results, we applied a simple median filter with a 3x3 kernel over each $p(z, \mu)$ plane. The results are shown in Fig. \ref{fig:detrates_lsst} as an average of all clusters.

\subsection{Volumetric supernova yield estimates}
\label{sec:volum_estimate}

The expected number of SNe of a subtype $j$, d$N_j$, in a comoving volume element d$V_c$, which is magnified by a factor $\mu$ and affected by extinction \textit{ext} at redshift $z$, depends on the volumetric SN rate $R^V_j$ and the survey control time, $T_j(z, \mu, ext)$, which is the total time during which the survey is sensitive to detecting an SN. The relationship is given by

\begin{equation}
\label{eq:number_element_2}
    \mathrm{d} N_j(z, \mu, ext) = T_j(z, \mu, ext) \frac{R^V_j}{1+z} \mathrm{d} V_c,
\end{equation}
where the $(1+z)$ factor corrects for cosmological time dilation. The comoving volume unit $\mathrm{d}V_c$ can be calculated as

\begin{equation}
    \mathrm{d} V_c = \frac{cd^2_L(z)}{H(z)(1+z)^2} \mathrm{d}\omega \mathrm{d}z,
\end{equation}
where d$\omega$ is the solid angle element for the survey, d$z$ is the redshift element, and $d_L$ is the cosmological luminosity distance for redshift $z$, and $H(z)$ is the Hubble parameter, as a function of redshift.

In the volumetric method, we integrated the expected number of detected SNe during a survey ${\rm d} N_j(z, \mu)$ in a comoving volume element ${\rm d} V_c$. This calculation depends on the survey control time $T_j(z, \mu, ext)$, and the volumetric SN rate $R^V_j(z)$, at redshift $z$, for an SN type $j$, assuming the volume element is magnified by a factor of $\mu$. We used survey control time estimates with a stochastic host extinction estimate, as described in Sect. \ref{sec:snana_hltds}.

To obtain a volumetric estimate of SN yields, we assumed volumetric CC SN rates, in units of ${\rm yr}^{-1} {\rm Mpc}^{-3} h_{70}^3$, of $R^V_{\rm CC}(z) = k_{\rm CC}^{V} h_{70}^2 \psi(z)$, where $k^V_{\rm CC} = 0.0091 \pm 0.0017 M_\odot^{-1}$ from \cite{2015ApJ...813...93S} is the fraction of stars that are CC SN progenitors, and $\psi(z)$ is the cosmic star formation history, using the updated values from \cite{2020ApJ...890..140S}, Table 2. Since $\psi$ scales with $h$, there is an additional $h^2$ factor in the relation. We extend this model to $z=6$. We assumed that the relative fractions of CC SN subtypes are constant throughout cosmic history and used the values proposed by \cite{2017PASP..129e4201S} with Type IIL and Type IIP relative rates from \cite{2011MNRAS.412.1441L}.

For SNe Ia, we used the simple volumetric rate provided by \cite{2020ApJ...890..140S}, which follows a broken power-law: $R^V_{\rm Ia} = R_0 (1+z)^A$, with $R_0 = 2.40 \pm 0.02 \times 10^{-5}~ {\rm yr^{-1}~ Mpc^{-3}}~ h^3_{70}$ and $A = 1.55 \pm 0.02$ for $z <= 1$, and $A = -0.1 \pm 0.2$ for $z > 1$.

For every cluster in the sample, we computed the expected yield of detected SNe as a function of redshift $\mathrm{d}N_j/\mathrm{d}z$, by integrating $\mathrm{d}N_j(z, \mu)$ given in Eq. \ref{eq:number_element_2} in the source plane at fixed redshifts, in bins of $\Delta z = 0.1$,

\begin{equation}
    \frac{\mathrm{d} N_j}{\mathrm{d}z} = R^V_j(z) \int \frac{c T_j(z, \mu, ext)d_L^2(z)}{H(z)(1+z)^3} \mathrm{d}\omega.
    \label{eq:volum_separate}
\end{equation}

\begin{figure*}[h]
\centering
    \centering
    \begin{subfigure}[b]{1.\columnwidth}
    \includegraphics[width=0.99\columnwidth]{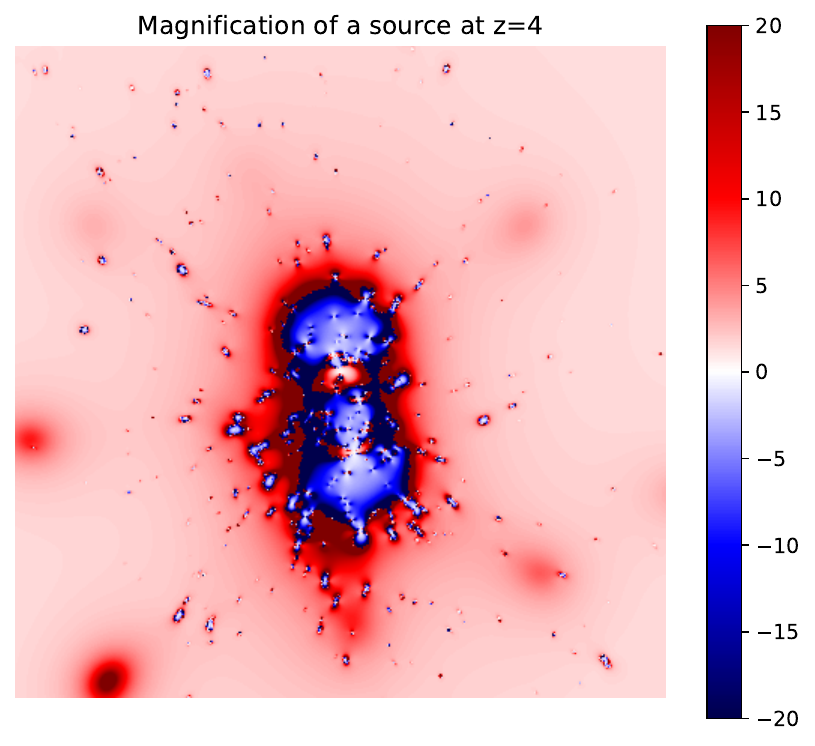}
    \end{subfigure}
    \begin{subfigure}[b]{0.9\columnwidth}
    \includegraphics[width=0.99\columnwidth]{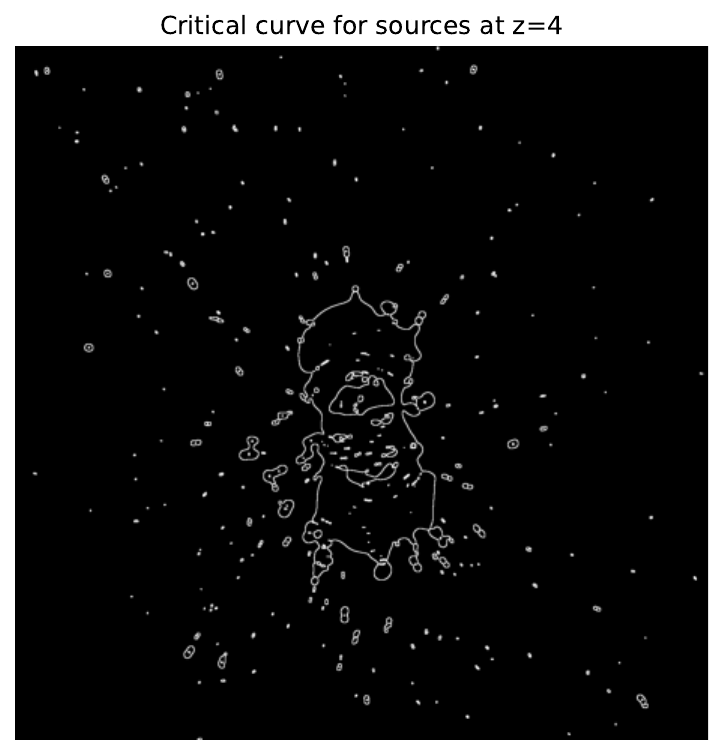}
    \end{subfigure}

    \begin{subfigure}[b]{0.9\columnwidth}
    \includegraphics[width=0.99\columnwidth]{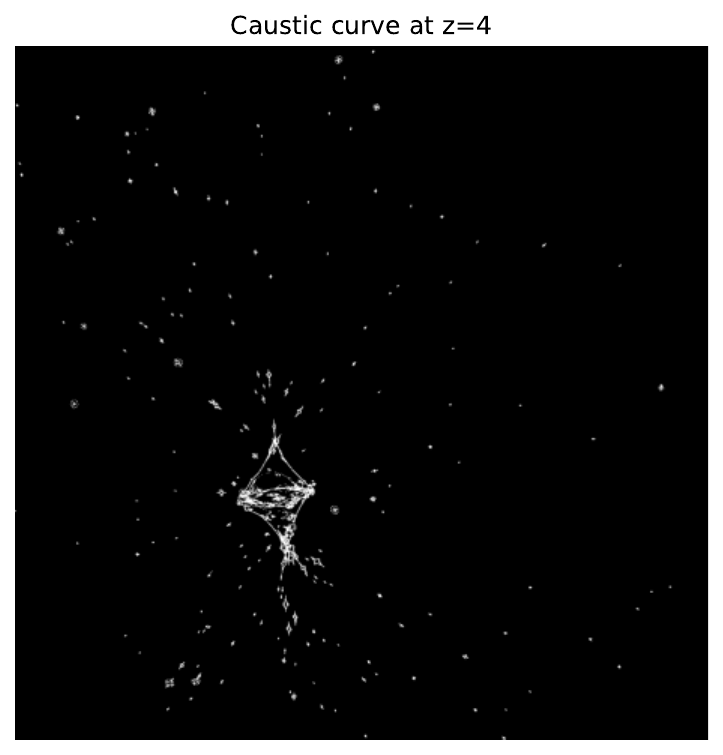}
    \end{subfigure}
    \hspace{0.1\columnwidth}
    \begin{subfigure}[b]{0.9\columnwidth}
    \includegraphics[width=0.99\columnwidth]{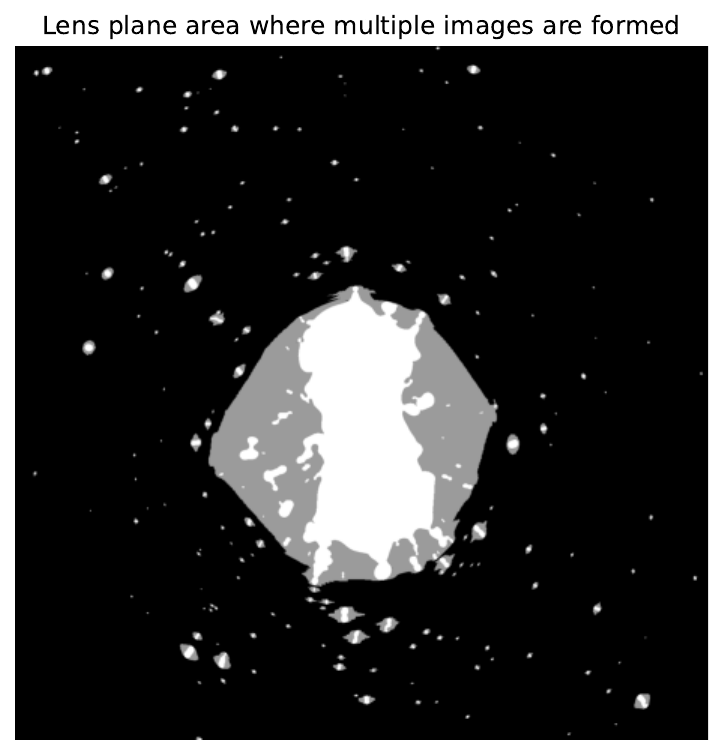}
    \end{subfigure}

\caption{Visualization of the volumetric estimate procedure. The plots are made based on model d1 FITS maps of Abell 370 by \cite{2023MNRAS.524.2883N}, trimmed to a central 1500-pixel by 1500-pixel square for clarity for the three lens plane plots, the bottom left is trimmed to 1000 by 1000 pixels. The source plane is fixed to $z=4$ in this example. 
Top left: magnification map for sources at redshift $z=4$. Negative values have the interpretation of a change in parity, namely, the image appears mirrored. To compute changes in flux, the absolute value is used.
Top right: The critical curve for a source plane at $z=4$. The critical curve is found by tracing the line at which magnification $\mu$ approaches $+\infty$ from one side, and $-\infty$ from the other. This line mapped to the source plane is the caustic curve, shown in bottom left.
Bottom left: Caustic curve for a source plane at $z = 4$. Sources within the caustic curve form multiple images, sources outside the caustic curve do not. 
Bottom right: lens plane area within the critical curve (white) and total area within which images belong to multiple image systems (white+gray). The latter can be obtained by tracing a ray for each pixel from the lens plane to the source plane, and checking whether the ray falls within or outside the caustic curve. This area is then used in our volumetric method, along with a magnification map, and a map of image multiplicity obtained through similar ray tracing method. We repeat this procedure for every source $z$ plane of interest.
}
\label{fig:A370_example}
\end{figure*}

To compute the volume multiply imaged by lensing clusters, we used lens models provided by authors listed in Table \ref{tab:cl_list}. We performed calculations using FITS file maps of convergence $\kappa$, shear $\gamma$ and deflection maps provided with lensing models where available. For models with no FITS file maps included, we generated high resolution maps using Lenstool\footnote{\url{https://projets.lam.fr/projects/lenstool/wiki}} \citep{2007NJPh....9..447J,2009MNRAS.395.1319J}. We then performed calculations using a Python toolkit\footnote{\url{https://github.com/mbronikowski/CLMIT}} developed for this work with the following procedure, which is visualized in Fig. \ref{fig:A370_example}. For each cluster, for each source redshift, we found the area inside the critical curve on the pixel map by calculating the magnification $\mu$ maps from shear and convergence, and finding the inside of the critical curve. We achieved this by finding the outermost area where $\mu < 0$, as well as all areas inside it. We then mapped that area to the source plane to find the area inside the caustic curve, and mapped that area back to the lens plane, both using deflection maps. This area is the total lens plane area in which images of sources belong to multiple image systems. Next, for each pixel in this lens plane area, we calculated the comoving volume to which it corresponds, and calculated the expected value of SNe discovered in that comoving volume, taking into account source redshift and magnification. Finally, we divided this value by the multiplicity of the image system the pixel belongs to, since otherwise each image of a given multiply imaged SN would be counted as a separate SN. We limited magnification to $\mu_{\rm max} = 100$ to mitigate numerical effects from finite-resolution cluster lensing maps. We also rejected demagnified images from the calculation. 

As volumetric SN rates at redshift $z>2.5$ are poorly constrained, one can expect them to be the dominant source of uncertainty of our results at high redshifts. This will improve with JWST targeted programs for observing samples of SNe at $z>2$  \citep[e.g.,][]{2024jwst.prop.5324P,2025ApJ...979..250D} and \textit{Roman}'s HLTDS, once it starts operating \citep{2018ApJ...867...23H}. We therefore provide separately an estimate of the integral from Eq. \ref{eq:volum_separate} $I_j(z) = \mathrm{d}N_j/\mathrm{d}z/R_j^V(z)$, in addition to the $\mathrm{d}N_j/\mathrm{d}z$ estimate. This value is independent of volumetric SN rates, and its uncertainty depends only on the uncertainty of the cluster lensing model and on the uncertainty of the probability of detecting an SN $p(z, \mu)$, the latter of which we assumed to be negligible. This value can be used to calculate a yield prediction which assumes a different volumetric SN rate $R'^V_j$, by simply multiplying the two values $dN'_j (z)= I_j(z)R'^V_j$. In the summary statistics in this work, we assumed that the uncertainty of $dN_j(z)$ is perfectly correlated between redshift bins within the same cluster, but uncorrelated between separate clusters.

To estimate the uncertainty of the value of the model-dependent component of the predicted SN yield estimate, $I_j(z)$, we performed the same volumetric calculations for a range of strong lensing models of the same galaxy cluster that have adopted different modelling algorithms and assumptions. We focus on the HFF galaxy cluster Abell 2744, which, owing to the extensive and high quality imaging and spectroscopic data, is one of the best studied lens clusters to date. We consider the following nine lens models resulting from the HFF program “CATS v4” \citep{2014MNRAS.444..268R}, ``Diego v4.1,'' ``GLAFIC v4'' \citep{2018ApJ...855....4K}, ``Keeton v4,'' ``Zitrin NFW v3,'' ``Zitrin LTM v4.1,'' ``Zitrin LTM Gauss v3,'' ``Sharon v4cor'' \citep{2014ApJ...797...48J}, and ``Williams v4.'' We also considered the lens model by \cite{2021AnA...646A..83R}, which exploits extensive MUSE observations of the galaxy cluster core. More recently, improved lens models of Abell 2744 have been developed by exploiting the large sample of new strong lensing features unveiled by the JWST NIRCam observations from the UNCOVER and GLASS-JWST surveys \citep{2023MNRAS.523.4568F}, in combination with the first MUSE observations of the so-called ’SL clump’ \citep{2023ApJ...952...84B}.  We also repeated the calculations for a set of MCMC chains which sample the probability distribution of solutions for the \cite{2023MNRAS.523.4568F} model to sample the uncertainty predicted by an individual, recent model.

With this approach, we obtained a conservative systematic uncertainty on the value of $I_j(z)$, as not all the lens models include large samples of securely identified multiple images (enabled by the most recent observations) or yield similar root-mean-square separation between the model-predicted and observed positions of the multiple images.

We discuss the results of these calculations in more detail in Sect. \ref{sec:error_budget}. We find that a flat $30\%$ uncertainty in $I_j(z)$ conservatively accounts for the systematic uncertainty inherent to the gravitational models.

\subsection{Observability constraints}

 Here, we consider the observability of the clusters in the different surveys. The HLTDS recommendations by \cite{2021arXiv211103081R} put constraints on the observable regions of the sky. The first condition is a high ecliptic latitude to minimize zodiacal light, and to reach the \textit{Roman} ``continuous-viewing zone'', specifically, $\gtrapprox 54^\circ$ from the ecliptic. This condition alone vastly limits the range of clusters available for HLTDS. Out of 71 clusters in the volumetric sample, only 8 fulfill this condition, with one additional cluster available if this constraint is relaxed to $\gtrapprox 50^\circ$. Out of those, only one cluster is within our arc-specific sample. 

This small sample is further reduced by the second constraint: a high galactic latitude to minimize dust extinction. The accessibility of individual clusters for HLTDS, as well as whether they fall within LSST's observing field, are listed in Table \ref{tab:cl_list}. As can be concluded from Fig. \ref{fig:extinction_roman}, there are only two clusters which fulfill both latitude conditions for HLTDS, or three if the ecliptic latitude condition is relaxed to $|\beta| \gtrapprox 50^\circ$: RXC J0232.2-4420, Abell S295 and, with the relaxed condition, Abell 1758a. 

The visibility of galaxy clusters to the \textit{Vera C. Rubin} Observatory is far greater. As the Observatory is going to observe a large area of the sky in its LSST, it will observe 45 of the clusters in our volumetric sample, 16 of which also belong to our arc-specific sample. These are listed in Table \ref{tab:cl_list}.

For completeness, we repeated the calculations performed in this work for every cluster in the sample, even if the cluster was not observable for a given survey, to offer a point of comparison for any proposed targeted cluster surveys. We have excluded those results from summary statistics, but we have made them available along with the rest of results.

\section{Results}
\label{sec:results}

\begin{figure*}[h]
\centering
    \centering
    \begin{subfigure}[b]{0.33\textwidth}
    \includegraphics[width=\textwidth]{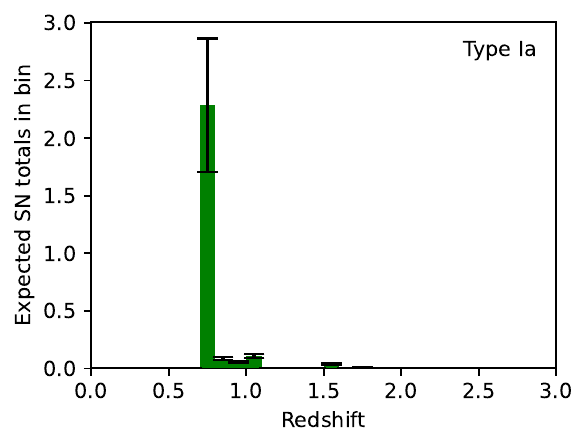}
    \end{subfigure}
    \begin{subfigure}[b]{0.33\textwidth}
    \includegraphics[width=\textwidth]{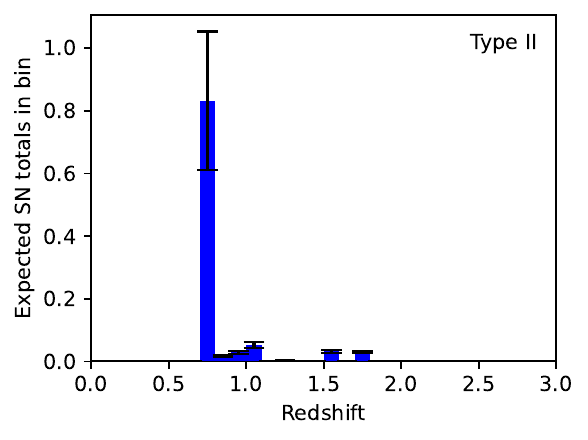}
    \end{subfigure}
    \begin{subfigure}[b]{0.33\textwidth}
    \includegraphics[width=\textwidth]{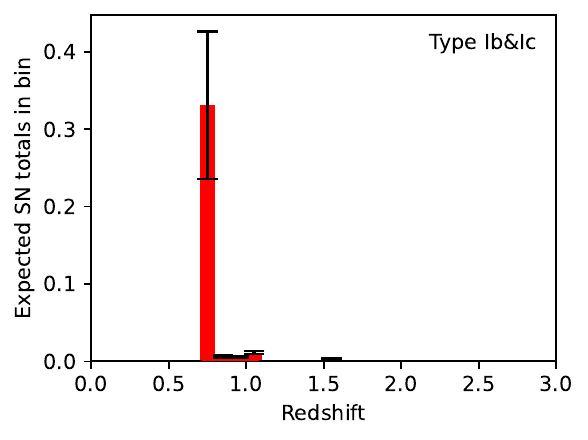}
    \end{subfigure}

\caption{Expected discovered multiply imaged SN distribution by type in LSST from the arc-specific method. The strong peak around $z \in [0.7, 0.9]$ can be attributed to a handful of  well-known massive star-forming galaxies in this redshift range, such as the grouping of galaxies lensed by Abell 370.}
\label{fig:specific_lsst}
\end{figure*}

\begin{figure*}[h]
\centering
    \centering
    \begin{subfigure}[b]{0.49\textwidth}
    \includegraphics[width=\textwidth]{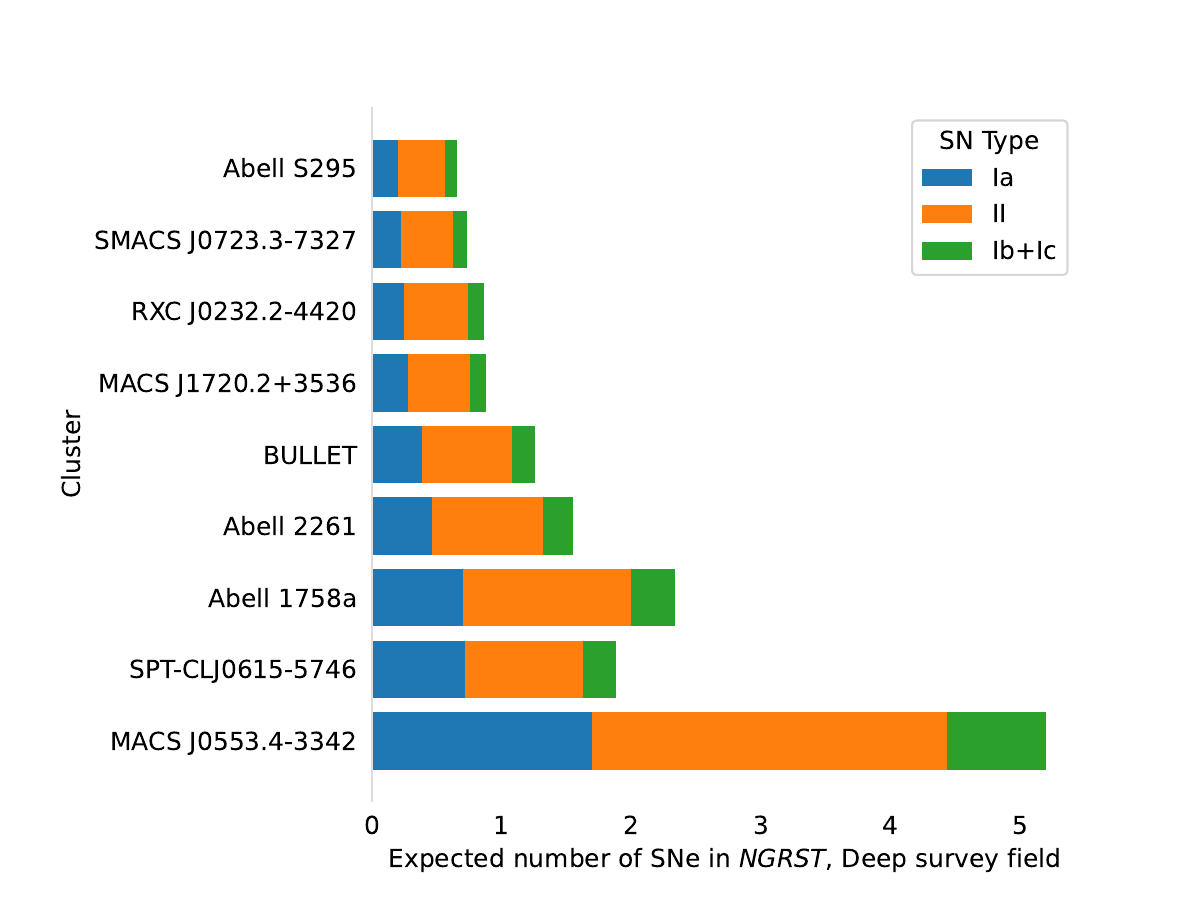}
    \end{subfigure}
    \begin{subfigure}[b]{0.49\textwidth}
    \includegraphics[width=\textwidth]{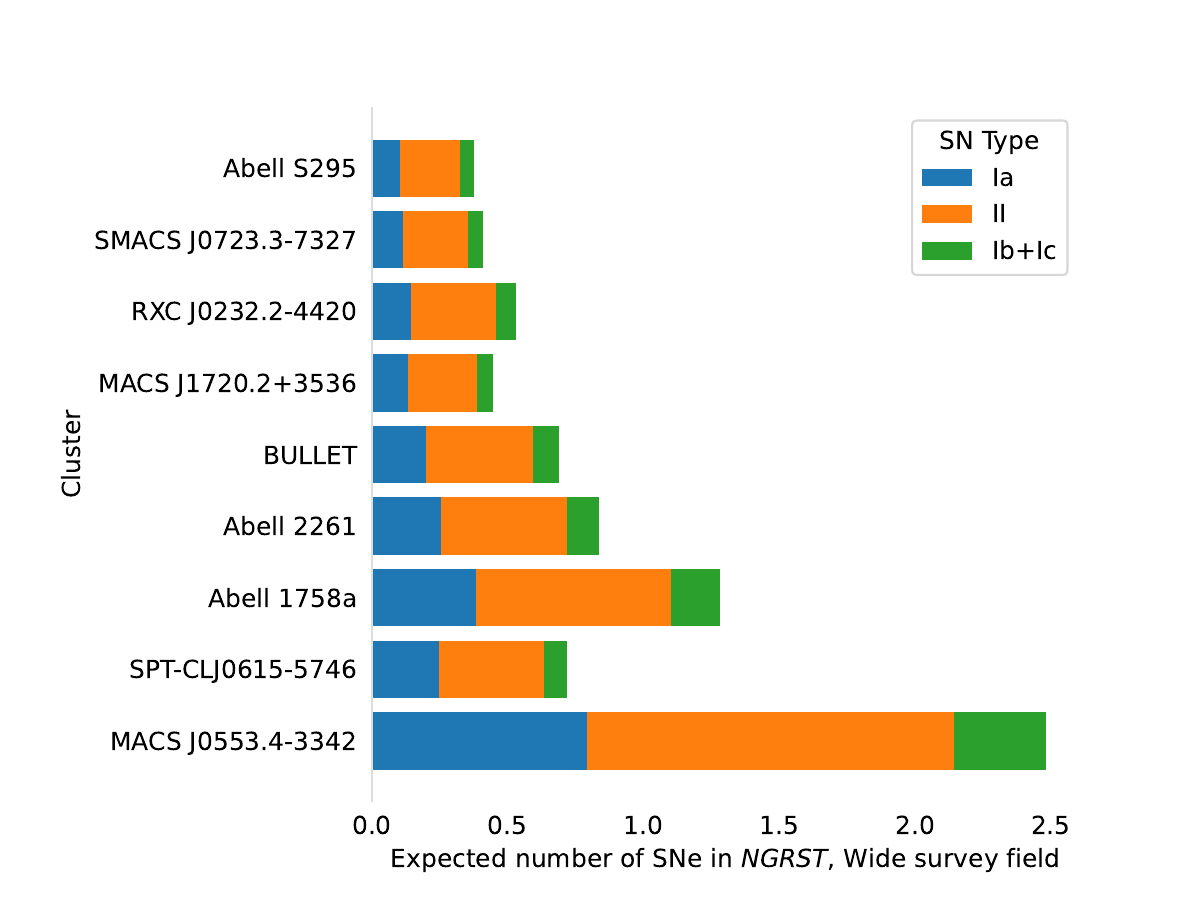}
    \end{subfigure}

\caption{Expected multiply imaged SN yields for the clusters estimated volumetrically, for the Deep (left) and Wide (right) survey tiers of HLTDS. For visual clarity, we plot total yields for Type IIL, IIP, IIb and IIn SNe, and Type Ib and Ic SNe, grouped together.}
\label{fig:observable_roman}
\end{figure*}

\begin{figure}[h]

    \includegraphics[width=0.95\columnwidth]{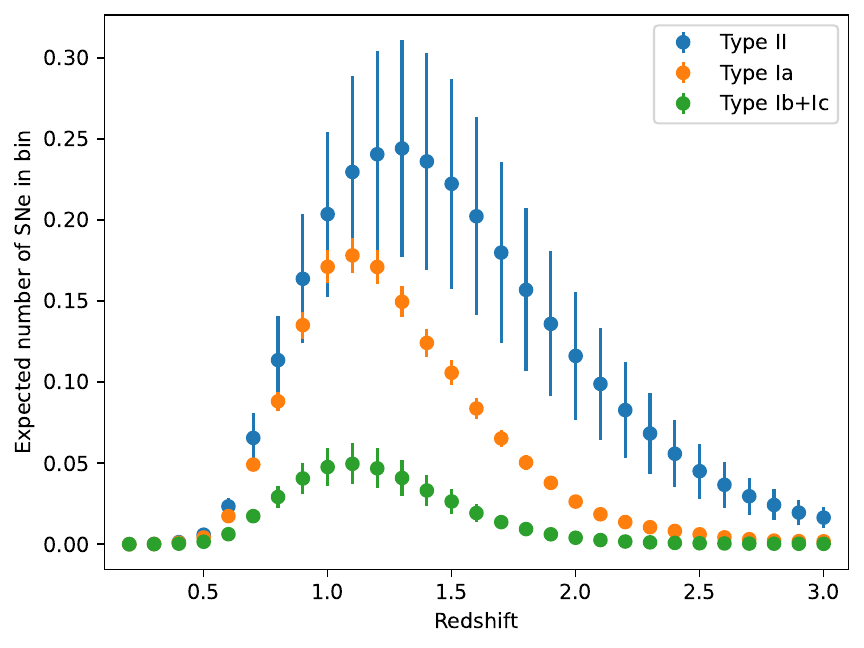}

\caption{Redshift distribution of the expected multiply imaged SN yields for the LSST baseline survey, estimated volumetrically, in all galaxy clusters. For visual clarity, we plot total yields for Types IIL, IIP, IIb, and IIn SNe, and Types Ib and Ic SNe, grouped together. Note: the $y$ axis is equal to $0.1 \cdot dN_{\rm j}/dz$.}
\label{fig:lsst_vol_z}
\end{figure}

\begin{figure}[h]

    \includegraphics[width=0.95\columnwidth]{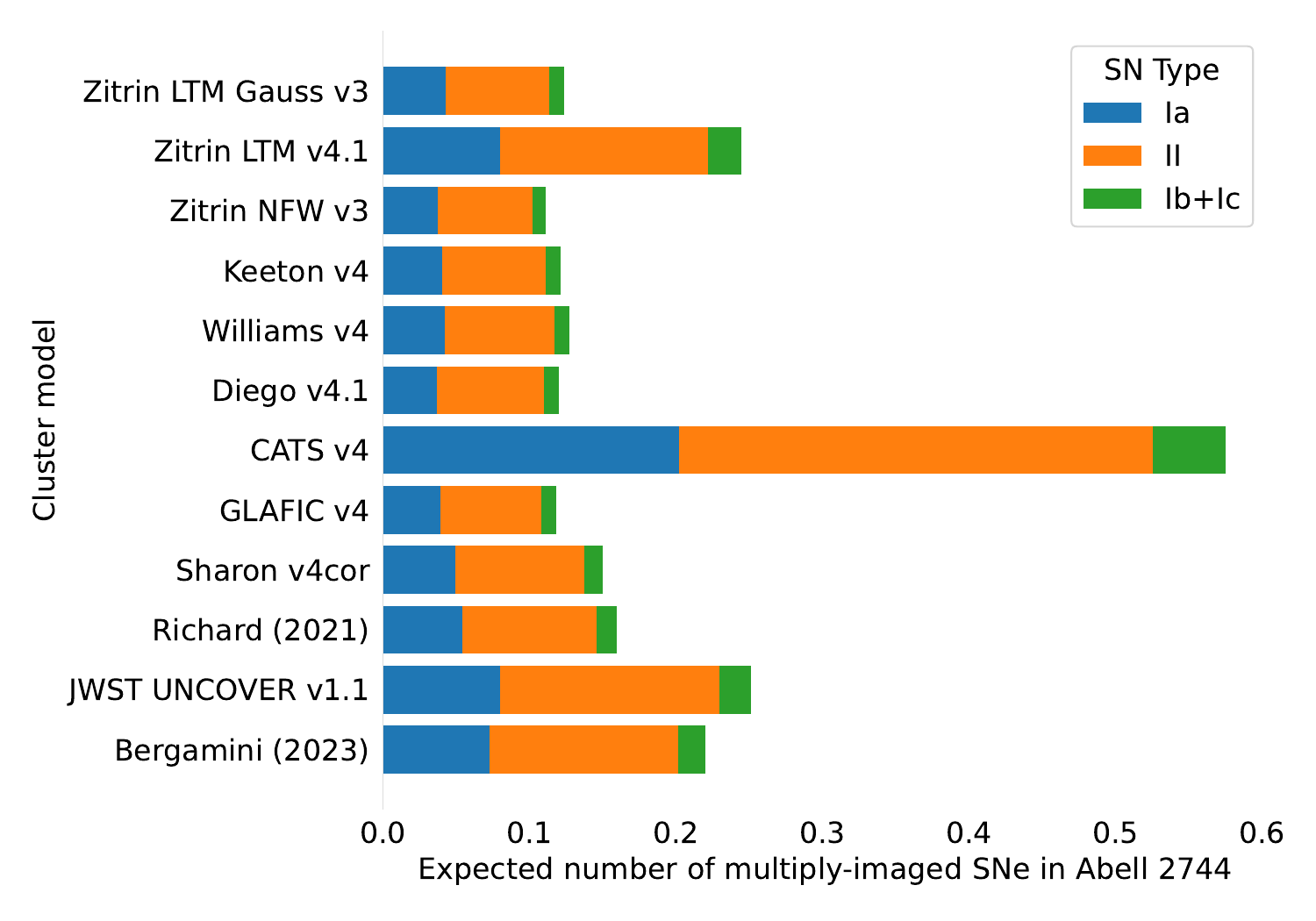}

\caption{Comparison between volumetric method SN yield predictions in LSST using lensing models from different sources, sorted chronologically by date of release. After rejecting the results from the CATS v4 model as a possible outlier, the relative uncertainty of the total prediction is around $30\%$ for all SN subtypes.}
\label{fig:model_comp}
\end{figure}

\begin{figure}[h]
\centering
    \centering
    \begin{subfigure}[b]{0.99\columnwidth}
    \includegraphics[width=0.99\columnwidth]{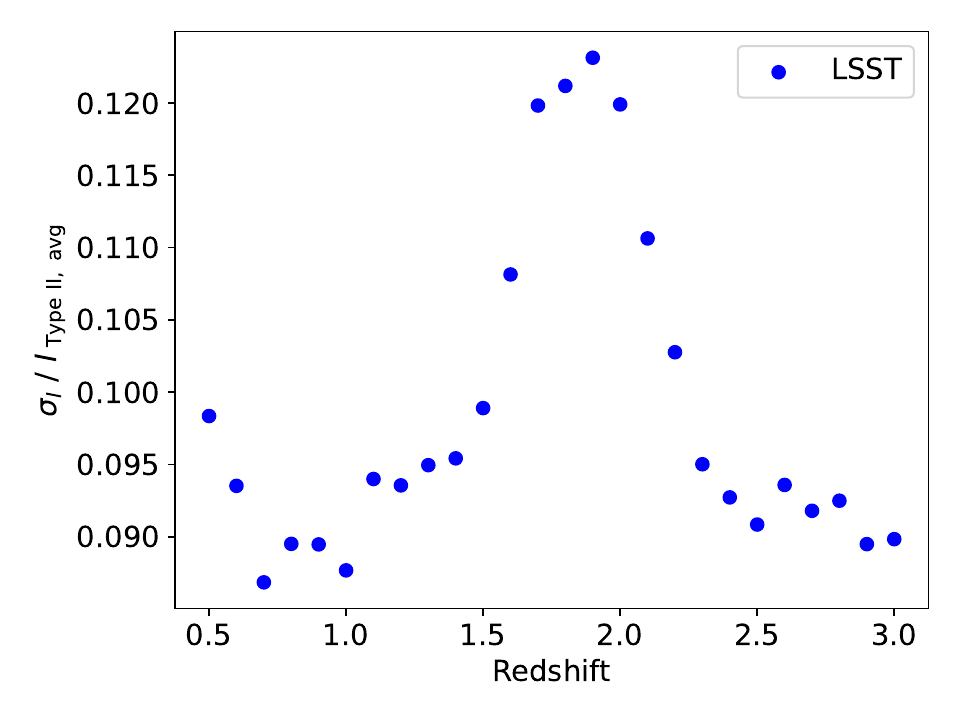}
    \end{subfigure}
    \begin{subfigure}[b]{0.99\columnwidth}
    \includegraphics[width=0.99\columnwidth]{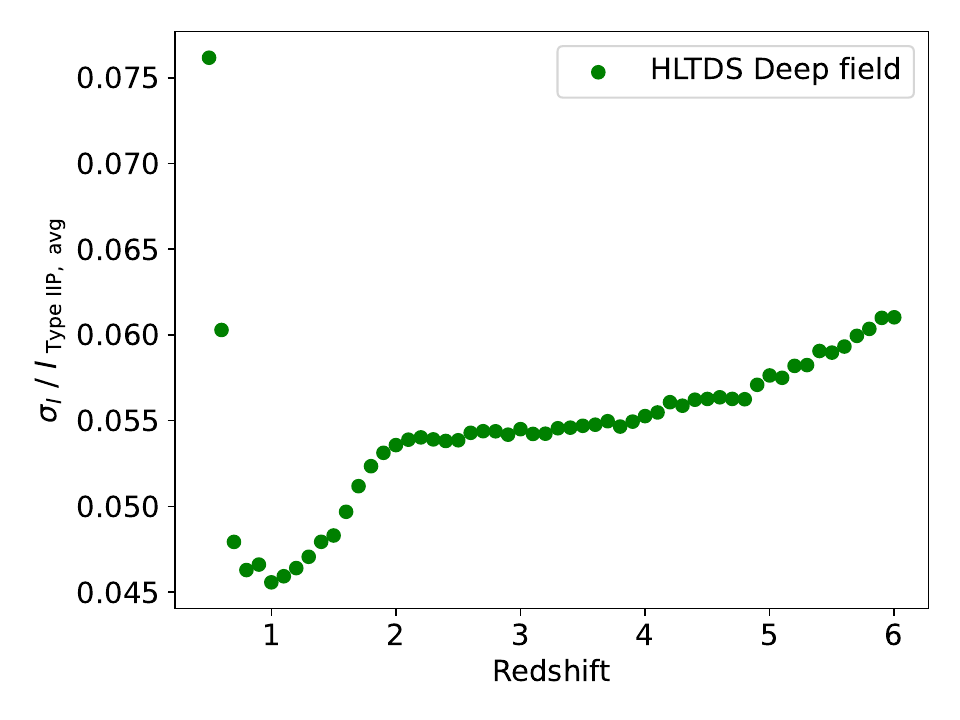}
    \end{subfigure}

\caption{Relative uncertainty estimates of the integral $I_j$, calculated as the standard deviation $\sigma_I$ of the value over 50 MCMC range models, divided by the average value $I_j$, for the UNCOVER survey model, for Type IIP + Type IIL SNe observed by LSST (top) and Type IIP SNe observed by HLTDS in the Deep survey field (bottom).}
\label{fig:uncover_error}
\end{figure}

In the sections below, we present the results of our estimates for the two surveys explored here,  HLTDS and LSST.

\subsection{Survey control time}

Figures \ref{fig:detrates_roman_deep}, \ref{fig:detrates_roman_wide}, and \ref{fig:detrates_lsst} show $p_j(z, \mu, ext)$ for all simulated SN subtypes for HLTDS Deep and Wide fields, and LSST, respectively -- the latter is an average for all clusters in the field. As those figures show, strong lensing enables the discovery of SNe which are normally too faint to be detected due to redshift. We found that \textit{Roman} will enable the discovery of SNe at redshifts up to $z \approx 6$ for certain subtypes, if they are sufficiently magnified. The sharp cutoff beyond a fixed redshift present in some of the results for HLTDS can be attributed to the SN spectrum being redshifted enough that the UV end of the model no longer covers the rest-frame wavelength of the reddest telescope filter.

\subsection{SN yields in the known arcs}

The SED fitting process yielded 312 arcs with good constraints on specific SN rates, including 286 observable to LSST, and 12 to HLTDS. 83\% of the systems we consider have three or fewer identified images in our sample, 15\% have four or five images, and only five systems, or $<2\%$ of our sample, have 6 or 7 images.

The yields in the arc-specific method depend on the galaxy’s SN rate $R_j$, and the survey control time $T_j$ (see Eq. \ref{eq:no_of_sne}).  The survey control time depends on redshift and magnification, as well as extinction parameters (see Eq. \ref{eq:surver_control_time}). The SN rate $R_j$, in turn, depends on the galaxy’s SFR, and the fraction of stars which explode as SNe $k_{\rm CC}$ (for CC SNe), or SFH and the conversion parameters $A$ and $B$ (for Type Ia). 

We propagated uncertainties from SFR,  $M_\star$, and the parameters $k_{\rm CC}$, $A$, and $B$, as these represent the primary sources of statistical uncertainty in our results. Notably, the uncertainties associated with SFR, $M_\star$, and the factors $k_{\rm CC}$ , $A$ and $B$ are of comparable magnitude. We found that the uncertainty contribution from magnification and redshift to be negligible, as we discuss in detail in Sect. \ref{sec:error_budget}. 

We note that the lensed images of the same system are simulated independently, without incorporating prior knowledge of the time delays between the images. To ensure a cautious estimate, we opted to average the number of expected multiply imaged SNe in an arc system across individual images, rather than summing them. This approach accounts for the fact that it is possible that an SN image could happen outside the survey duration (e.g., in the case when only the last SN image is detected). By averaging instead of summing, we avoid overestimating the number of detectable events while maintaining a balanced interpretation of the results.

\subsubsection{Arc-specific predictions for HLTDS}

Abell S295 is the only cluster observable by HLTDS which falls in our photometric sample. We estimate that if the cluster is observed, HLTDS will discover $0.341 \pm 0.028$ Type Ia SNe, as well as $ 0.1194 \pm 0.0085 $ Type Ib and Ic SNe, and $0.477 \pm 0.036$ Type II SNe. Notably, these values virtually do not depend on the survey tier the cluster is observed in, since the arcs for which we have reliable SFR and $M_\star$ constraints are all placed at relatively low redshifts $z < 2.2$, where the probability of SN detection in the HLTDS is very close to 1, as can be seen in Figs. \ref{fig:detrates_roman_deep} and \ref{fig:detrates_roman_wide}.

\subsubsection{Arc-specific predictions for LSST}

Figure \ref{fig:specific_lsst} presents predicted numbers of multiply imaged SNe in LSST according to the arc-specific estimate method. The distribution shows that most of the SNe are expected to be discovered around redshift $z = 0.8$. Using this method, we predict that LSST will discover $1.00 \pm 0.22$ Type II SNe, $ 2.59 \pm 0.58 $ Type Ia SNe and $ 0.359 \pm 0.095$ Type Ib and Ic SNe which are multiply imaged.

\subsection{SN yields in the volume probed behind the clusters}
\label{sec:res_volum}

\subsubsection{Volumetric predictions for HLTDS}

As outlined earlier, HLTDS will only be able to observe a handful of clusters from our sample. Specifically, following \cite{2021arXiv211103081R}, only 8 galaxy clusters meet the HLTDS requirement for continuous viewing fields of ecliptic latitude $|\beta| > 54^\circ$. If this condition can be relaxed, even by a few degrees, Abell 1758a also becomes observable. Within this sample of 9 clusters, only three have a high galactic latitude $|b| > 55^\circ$: Abell 1758a, RXC J0232.2-4420 and Abell S295.

Figure \ref{fig:observable_roman} and Table \ref{tab:vol_roman_percl} show volumetric SN yield predictions for \textit{Roman}, assuming that a given cluster falls within one of the observing fields of either tier of HLTDS, for all 9 clusters that meet (or nearly meet) the ecliptic latitude constraint. Only Abell S295 and RXC J0232.2-4420 strictly meet HLTDS's requirements and these two clusters are predicted to yield the fewest multiply imaged SNe. Far better results can be achieved if the requirements are relaxed. If the $|\beta| > 54^\circ$ requirement is relaxed, Abell 1758a can be observed by the HLTDS, which would enable the discovery of around two multiply imaged SNe. On the other hand, if fields with lower galactic latitude are permitted, then promising galaxy clusters such as SMACS J0723.3-7327, SPT-CLJ0615-5746, and MACS J0553.4-3342 become available. The latter cluster is of particular interest, as we predict that \textit{Roman} can discover 2-5 multiply imaged SNe in the field, depending on which survey tier the cluster falls in. 

\subsubsection{Volumetric predictions for LSST}

While the LSST survey strategy is not fully decided upon, it is currently known what declination range is going to be covered, and thus which galaxy clusters are observable by LSST. Assuming the baseline survey strategy, we calculated the estimated yields from LSST in the first three years of its operations. To calculate the expected values for the full 10-year survey, a sufficiently accurate approximation is to multiply the values by $10/3$. It is important to note, however, that this is an approximation, and not an exact value, because of ``edge-case'' SNe, which may explode before the survey begins, or peak after the survey ends, but can still be detected.

Overall, we expect LSST to discover $3.02\pm0.86$ Type II SNe, $1.528\pm0.055$ Type Ia SNe and $0.40\pm0.10$ Type Ib/Ic SNe in its first three years of operation in clusters within our sample. The redshift distribution of multiply imaged SNe expected to be discovered by LSST is shown in Fig. \ref{fig:lsst_vol_z}.  Remarkably, we predict that LSST should be able to detect Type II SNe at relatively high redshifts, provided that they are sufficiently magnified -- our prediction puts the expected number of Type II SNe at $z \geq 2.5$ at $0.171\pm0.066$, and Type Ia SNe at $0.0193\pm0.0024$. Thus, we expect a greater than 50\% chance of LSST detecting a $z \geq 2.5$ SN in its entire 10-year duration. The predictions for each individual cluster are shown in Table \ref{tab:vol_lsst_percl}.

\section{Discussion} \label{sec:discussion}

\subsection{Error budget}
\label{sec:error_budget}

Here, we discuss the sources of uncertainty in our calculated yields. In the arc-specific method, we constrained SFHs of arcs using their available photometry, by assuming a specific set of SFH models to which we fit the photometric data. There is a possible source of systematic uncertainty, as some of the arcs may have unusual SFHs which produce similar SEDs as those we assumed. Furthermore, photometric catalogues provided by ALCS, which we used for clusters other than those studied by \cite{2021AnA...646A..83R}, or Abell 2744, do not take the complex shapes of some arc images into account. This may result in parts of the image being excluded from the photometric measurement, and therefore an underestimated flux. Lastly, we assumed fixed values for redshift and magnification of arcs provided by catalogues, and neglected the uncertainty of magnification, and the uncertainty of redshift for arcs with no spectroscopic redshift constraints. This is done in part to simplify SED fitting to galaxies and in part due to the fact that these estimates are not consistently provided by lens model authors. Where redshift uncertainties are available, however, we find that the median redshift uncertainty is sufficiently small, around $\Delta z < 0.2$, that it should not be a significant contributor of uncertainty. Similarly, the average magnification uncertainty is expected to be small, based on clusters where it is provided. For example, the average relative magnification uncertainty provided by \cite{2021AnA...646A..83R} for lensed images behind the cluster MACS J1206.2-0847 is $1.7\%$. While we expect an additional systematic uncertainty in magnification estimates, as described in the following paragraph, estimating it for the arc-specific method is beyond the scope of this paper. We note, however, that the requirement of high quality broadband photometric and spectroscopic data for the arc-specific method means that the lens models are of higher average quality than that of the broader sample for the volumetric method. This implies a lower uncertainty on magnification and model-derived redshift. We note that the uncertainties of redshift and magnification of arc in the literature, are not always consistently provided by authors.

For the volumetric method, we assumed a 30\% uncertainty for all redshifts on the integral $I_j(z)$ corresponding to all parameters of the volumetric estimate calculations other than the volumetric SN rate. This uncertainty dominates the estimates yield at redshifts below $z < 3$. At higher redshifts, however, the uncertainty is dominated by the fact that volumetric SN rates are poorly constrained due to the rarity of high-redshift SN detections. We believe 30\% is justified to account not only for model-specific uncertainties, but also all the systematic errors, including previously undiscovered clumps, as was the case with older models of Abell 2744. This can be seen in Fig. \ref{fig:model_comp}, where \cite{2021AnA...646A..83R} suspected a previously undiscovered clump but did not fully model it, but then models by \cite{2024ApJS..275...36F} and \cite{2023ApJ...952...84B} incorporate it. Furthermore, systematic uncertainty is exacerbated by all the assumptions that are typically made in cluster modeling. We note that due to limited availability of newer models, some of the models we used are very old -- nearly 10 years at the time of writing -- which further justifies a high assumed uncertainty.

To show the uncertainties that are typical for a recent model which incorporates high quality JWST data, we estimated the relative uncertainties of $I_j(z)$ from MCMC chains provided for the UNCOVER survey model of Abell 2744 by \cite{2024ApJS..275...36F}. Figure \ref{fig:uncover_error} shows those uncertainties, defined as the standard deviation of $I_j$ normalized by the average, for Type IIP and IIL SNe for LSST, and for Type IIP SNe for HLTDS. As can be clearly seen, the statistical uncertainty is around $5.5\%$ for HLTDS, and $10\%$ for LSST on average. We believe, however, that this is only because this is a model which incorporates state-of-the-art data, which is not true for our entire volumetric sample.

\subsection{Differences between predictions of the employed methods}

As the two methods applied in our work are sensitive to different aspects of galaxy clusters, it is expected that the results will differ as well. The volumetric method is only sensitive to the lensing properties of a cluster and agnostic of actual galaxy distributions in  the lensed volume in which sources are multiply imaged; on the other hand, the arc-specific approach is only sensitive to bright galaxies with high quality broadband photometry, and is naturally less sensitive to high-redshift sources, due to the high luminosity distance, which effects low quality of photometry. 

Indeed, there is a noteworthy difference between the results of both multiply imaged SN yield estimate methods for LSST. The arc-specific method predicts $3.95 \pm 0.89$ to be detected in the first three years of the survey. On the other hand, if the cluster sample for the volumetric estimate is limited to only the same clusters as in the arc-specific estimate, the total number of expected SNe is $1.83\pm0.37$. On the other hand, the redshift distribution predicted by the volumetric method peaks around redshifts $z\in[1.0, 1.3]$ depending on the SN type, while the arc-specific expected distribution peaks around $z = 0.8$; the former tentatively appears to be more in line with the multiply imaged SNe discovered so far.

We believe that the difference in expected yields can, at least partially, be explained by the selection bias in our cluster sample. Specifically, we selected clusters which were studied in depth for their gravitational properties, the possibility of which is dependent on the presence of bright multiply imaged galaxies behind the clusters. Therefore, our selection criteria favor clusters with overdensities of multiply imaged bright galaxies behind them, whereas our volumetric method assumes a uniform volumetric SN rate that is agnostic of local matter density. On the other hand, the volumetric method accounts for galaxies which may be present behind the clusters but are too faint or distant to be used as a constraint in cluster modeling. We believe that the volumetric method's results can be used as lower limit constraints on the expected numbers of multiply imaged SNe, especially at lower redshifts, where the aforementioned overdensities are typically present.

It is important to note that we only provide predictions for a sample of largest, most well-studied clusters. While HLTDS is going to observe only a relatively small, carefully selected field, LSST is going to observe a large fraction of the sky, which will include numerous galaxy clusters which do not belong to our sample, the majority of which will be clusters with smaller Einstein radii, and thus have fewer arc systems behind each cluster \citep[e.g.,][]{2009arXiv0912.0201L}. Given that the majority of arcs are expected to be found behind smaller clusters, which fall outside this work's cluster sample, one can expect that most cluster-lensed SNe will be discovered behind small clusters. However, the SNe may be less useful for time-delay cosmography due to limited model constraint availability in small clusters with few arcs. A more thorough investigation of prospects for small and previously undiscovered clusters is a valuable topic of future research.

\section{Conclusions} \label{sec:conclusion}

In this paper, we estimate the expected number of multiply imaged SNe in well-studied clusters by the upcoming surveys by  \textit{Vera C. Rubin} Observatory's LSST and  \textit{Nancy Grace Roman} Space Telescope's HLTDS. We employed two approaches: one based on the expected SN rates in the known arcs behind galaxy clusters and another based on the volume probed behind the clusters. We found 285 arcs for which it was possible to employ the first method, 12 (143), of which are observable by the HLTDS (LSST). For the volumetric method, we found 71 clusters from the literature, 9 (46) of which are observable by the HLTDS (LSST). Here are our main conclusions:
\begin{itemize}
\item LSST offers great prospects for discovering multiply imaged SNe. LSST offers the capability to discover  $4.94\pm 1.02$ (from the volumetric method) or $3.95 \pm 0.89$ (from the arc-specific method) multiply imaged SNe in the first 3 years of the survey. This assumes that LSST follows the baseline survey strategy in that period. If the survey strategy is changed, for instance, to a rolling cadence one, where certain fields will be observed more often than others, the impact to the expected lensed SNe could be significant.

\item In the arc-specific approach, there is only one cluster that falls within \textit{Roman}'s observable fields, Abell S295. The expectations for Abell S295 are $0.341 \pm 0.028$ Type Ia SNe, $ 0.1194 \pm 0.0085 $ Type Ib and Ic SNe, and $ 0.477 \pm 0.036 $ Type II SNe.

\item We also calculated volumetric estimates of multiply imaged SN prospects for nine clusters that fulfill the HLTDS's requirement on high ecliptic latitude. We found that the galaxy cluster MACS 0553.4-3342 provides the highest number of lensed SN discoveries, at $5.2 \pm 2.2$ or $2.5 \pm 1.0$, depending on whether the cluster is observed in the Deep or Wide survey tier. On the other hand, if the cluster's region is rejected as a possible observing field  due to galactic latitude constraints, we find that the relaxation of the ecliptic latitude condition $|\beta| < 54^\circ$ to $|\beta| < 50^\circ$ unlocks the availability of another cluster: Abell 1758a, which would also result in the discovery of $2.34 \pm 0.97$ or $1.28 \pm 0.51$ multiply imaged SNe, if observed. 

\item If the aforementioned clusters are not observed in HLTDS, we recommend for another survey to be proposed to target them specifically, due to the high expected yield of multiply imaged SNe. Furthermore, we provide raw results related to the theoretical yields for clusters that fall outside of the HLTDS observing field constraints to facilitate the planning of similar surveys.

\item Cluster-lensed SNe can complement the scientific yield from galaxy-lensed SNe because they are susceptible to different systematics and are characterized by longer time delays. For this purpose, we recommend performing surveys that target promising, massive galaxy clusters at high depths, as well as follow-up campaigns to accurately measure the light curves of detected lensed SNe.

\end{itemize}

\section*{Data availability}

All data products associated with this paper are publicly available at \url{https://github.com/mbronikowski/Bronikowski25} or via Zenodo: \url{https://doi.org/10.5281/zenodo.15115001}.

\begin{acknowledgements}
MB and TP acknowledge the financial support from the Slovenian Research Agency (grants I0-0033, P1-0031, J1-8136, J1-2460 and Z1-1853) and the Young Researchers program. 
This work was also supported with travel grants by the Slovenian Research Agency (BI-US/24-26-085, BI-VB/23-25-005 and 
BI-US/22-24-006) and the COST Action CA21136 Cosmoverse. This research was supported by the Munich Institute for Astro-, Particle and BioPhysics (MIAPbP), which is funded by the Deutsche Forschungsgemeinschaft (DFG, German Research Foundation) under Germany's Excellence Strategy – EXC-2094 – 390783311.
AA acknowledges financial support through the project PID2022-138896NB-C51 (MCIU/AEI/MINECO/FEDER, UE) Ministerio de Ciencia, Investigaci\'on y Universidades.
This work is based on observations taken by the RELICS Treasury Program (GO 14096) with the NASA/ESA HST, which is operated by the Association of Universities for Research in Astronomy, Inc., under NASA contract NAS5-26555. 
This work utilizes gravitational lensing models produced by PIs Bradač, Natarajan \& Kneib (CATS), Merten \& Zitrin, Sharon, Williams, Keeton, Bernstein and Diego, and the GLAFIC group. This lens modeling was partially funded by the HST Frontier Fields program conducted by STScI. STScI is operated by the Association of Universities for Research in Astronomy, Inc. under NASA contract NAS 5-26555. The lens models were obtained from the Mikulski Archive for Space Telescopes (MAST).
N. B. acknowledges to be funded by the European Union (ERC, CET-3PO, 101042610). Views and opinions expressed are however those of the author(s) only and do not necessarily reflect those of the European Union or the European Research Council Executive Agency. Neither the European Union nor the granting authority can be held responsible for them.
We are grateful for the support of the University of Chicago’s Research Computing Center for assistance with the calculations carried out in this work.
We thank G. Caminha and L. Strolger for the useful discussions.
We thank the anonymous referee, whose valuable insight greatly improved our work.

The following software was used in this work: 
    SNCosmo \citep{barbary_2025_14714968},
    extinction \citep{barbary_2016_804967}, 
    NumPy \citep{harris2020array}, 
    Astropy \citep{2022ApJ...935..167A}, 
    CIGALE \citep{2019A&A...622A.103B}, based on \cite{2009A&A...507.1793N} and \cite{2005MNRAS.360.1413B}, 
    dustmaps \citep{2018JOSS....3..695M}, 
    Lenstool \citep{2007NJPh....9..447J,2009MNRAS.395.1319J},
    Matplotlib \citep{Hunter:2007}, 
    SNANA \citep{2009PASP..121.1028K}, 
    TOPCAT \citep{2005ASPC..347...29T}.
    
\end{acknowledgements}

\bibliography{bibliography}

\appendix

\onecolumn

\begin{landscape}
\section{Tables}

\begin{longtable}{lrrrrrcccl}

\caption{Galaxy clusters considered in this work, with their observability conditions for HLTDS and LSST, and sources of lensing models.} \\
\hline 
\hline
\label{tab:cl_list}

Cluster& z & RA $[^{\circ}]$ & DEC $[^{\circ}]$ & $\beta$ $[^{\circ}]$ & $b$ $[^{\circ}]$ & $|\beta| > 54^\circ$? & LSST? & Phot. sample & source \& comments \\ 
\hline
\endfirsthead
\caption{Continued.}\\
\hline \hline
Cluster& z & RA $[^{\circ}]$ & DEC $[^{\circ}]$ & $\beta$ $[^{\circ}]$ & $b$ $[^{\circ}]$ & $|\beta| > 54^\circ$? & LSST? & Phot. sample & source \& comments \\ 
\hline
\endhead
\hline
\endfoot

Abell 1758a & 0.280 & 203.186 & 50.543 & 53.893 & 65.297 &  &  &  & A, B \\
Abell 1763 & 0.228 & 203.834 & 41.001 & 46.323 & 73.445 &  &  &  & A \\
Abell 2163 & 0.203 & 243.954 & -6.145 & 14.899 & 30.459 &  &  & Y & C \\
Abell 2537 & 0.297 & 347.093 & -2.192 & 3.078 & -54.920 &  & Y & Y & C \\
Abell 2813 & 0.292 & 10.853 & -20.628 & -23.162 & -83.246 &  & Y &  & A \\
Abell 3192 & 0.425 & 59.725 & -29.925 & -49.063 & -49.012 &  & Y &  & A \\
Abell 697 & 0.282 & 130.740 & 36.367 & 17.538 & 37.265 &  &  &  & D \\
ACT-CLJ0102-49151 & 0.870 & 15.734 & -49.268 & -49.963 & -67.750 &  & Y & Y & \cite{2023AnA...678A...3C} \\
Abell S295 & 0.300 & 41.373 & -53.041 & -63.025 & -56.570 & Y & Y & Y & D \\
CLJ0152.7-1357 & 0.833 & 28.178 & -13.955 & -23.801 & -70.561 &  & Y &  & \cite{2019ApJ...874..132A} \\
MACS J0025.4-1222 & 0.586 & 6.388 & -12.388 & -13.890 & -74.060 &  & Y &  & A \\
MACS J0035.4-2015 & 0.352 & 8.859 & -20.262 & -22.037 & -82.219 &  & Y &  & A \\
MACS J0159.8-0849 & 0.405 & 29.955 & -8.833 & -19.703 & -65.587 &  & Y & Y & A \\
MACS J0257.1-2325 & 0.505 & 44.286 & -23.435 & -38.296 & -61.418 &  & Y &  & A \\
MACS J0308.9+2645 & 0.356 & 47.233 & 26.761 & 8.764 & -26.781 &  &  &  & E \\
MACS J0417.5-1154 & 0.443 & 64.394 & -11.909 & -32.705 & -39.478 &  & Y & Y & \cite{2019ApJ...873...96M} \\
MACS J0553.4-3342 & 0.430 & 88.354 & -33.711 & -57.135 & -26.009 & Y & Y &  & A \\
MS1008.1-1224 & 0.306 & 152.635 & -12.665 & -22.306 & 34.265 &  & Y &  & A \\
PLCK G171.9-40.7 & 0.270 & 48.237 & 8.372 & -9.204 & -40.657 &  &  & Y & E \\
PLCK G209.79+10.23 & 0.677 & 110.599 & 7.409 & -14.533 & 10.233 &  & Y &  & A \\
PLCK G287.0+32.9 & 0.390 & 177.709 & -28.082 & -26.482 & 32.907 &  & Y &  & \cite{2017ApJ...839L..11Z} \\
RXC J0018.5+1626 & 0.546 & 4.640 & 16.438 & 13.224 & -45.710 &  &  &  & A \\
RXC J0032.1+1808 & 0.396 & 8.046 & 18.133 & 13.451 & -44.499 &  &  & Y & \cite{2020ApJ...898....6A} \\
RXC J0142.9+4438 & 0.341 & 25.730 & 44.635 & 31.448 & -17.280 &  &  &  & C \\
RXC J0232.2-4420 & 0.284 & 38.077 & -44.347 & -54.761 & -63.448 & Y & Y &  & A \\
RXC J0600.1-2007 & 0.460 & 90.034 & -20.136 & -43.575 & -20.027 &  & Y &  & A \\
RXC J0911.1+1746 & 0.505 & 137.798 & 17.775 & 1.469 & 38.650 &  &  &  & A \\
RXC J0949.8+1707 & 0.383 & 147.466 & 17.119 & 3.763 & 47.014 &  &  &  & A \\
RXC J2211.7-0350 & 0.397 & 332.941 & -3.829 & 6.851 & -45.358 &  & Y & Y & C \\
1RXS J060313.4+421231 N & 0.228 & 90.820 & 42.245 & 18.807 & 9.737 &  &  &  & A, F \\
1RXS J060313.4+421231 S & 0.228 & 90.851 & 42.159 & 18.721 & 9.718 &  &  &  & A, F \\
SMACS J0723.3-7327 & 0.390 & 110.828 & -73.454 & -80.187 & -23.701 & Y &  &  & \cite{2022AnA...666L...9C} \\
SPT-CLJ0615-5746 & 0.972 & 93.966 & -57.780 & -81.031 & -27.331 & Y & Y &  & \cite{2018ApJ...863..154P} \\
WHL J013719.8-082841 & 0.566 & 24.354 & -8.456 & -17.288 & -68.390 &  & Y &  & A \\
Abell 370 & 0.375 & 39.971 & -1.582 & -16.306 & -53.560 &  & Y & Y & \cite{2023MNRAS.524.2883N} model d1 \\
Abell 2744 & 0.308 & 3.586 & -30.400 & -29.061 & -81.241 &  & Y & Y & \cite{2023ApJ...952...84B} \\
Abell S1063 & 0.348 & 342.183 & -44.531 & -33.825 & -59.933 &  & Y & Y & \cite{2024MNRAS.527.3246B} \\
1E 0657-56 (Bullet) & 0.296 & 104.659 & -55.957 & -77.337 & -21.235 & Y & Y &  & G \\
MACS J0257.6-2209 & 0.322 & 44.421 & -22.155 & -37.146 & -60.963 &  & Y &  & G \\
MACS J0329.6-0211 & 0.450 & 52.423 & -2.196 & -20.498 & -44.674 &  & Y &  & G \\
MACS J0416.1-2403  & 0.397 & 64.038 & -24.067 & -44.484 & -44.058 &  & Y & Y & \cite{2023AnA...674A..79B} \\
MACS J0451.9+0006 & 0.550 & 72.978 & 0.105 & -22.251 & -26.268 &  & Y &  & H \\
MACS J0520.7-1328 & 0.336 & 80.175 & -13.480 & -36.514 & -26.108 &  & Y &  & H \\
MACS J0940.9+0744 & 0.335 & 145.224 & 7.740 & -5.811 & 40.970 &  & Y &  & G \\
MACS J1206.2-0847  & 0.438 & 181.551 & -8.801 & -7.454 & 52.436 &  & Y & Y & G \\
MACS J2214.9-1359 & 0.502 & 333.739 & -14.004 & -2.937 & -51.287 &  & Y &  & G \\
RX J1347.5-1145 & 0.451 & 206.878 & -11.753 & -0.620 & 48.808 &  & Y & Y & G \\
SMACS J2031.8-4036 & 0.331 & 307.972 & -40.625 & -21.062 & -35.702 &  & Y &  & G \\
SMACS J2131.1-4019 & 0.442 & 322.770 & -40.322 & -24.218 & -46.921 &  & Y &  & G \\
SMACS J2332.4-5358 & 0.398 & 353.114 & -53.974 & -45.559 & -59.479 &  & Y &  & I \\
Abell 1423 & 0.213 & 179.322 & 33.611 & 30.262 & 76.659 &  &  &  & J \\
Abell 209 & 0.206 & 22.969 & -13.611 & -21.517 & -73.510 &  & Y &  & J \\
Abell 2261 & 0.224 & 260.613 & 32.133 & 55.117 & 31.861 & Y &  &  & J \\
Abell 383 & 0.187 & 42.011 & -3.533 & -18.799 & -53.502 &  & Y & Y & J \\
Abell 611 & 0.288 & 120.237 & 36.057 & 15.200 & 28.900 &  &  &  & J \\
CLJ1226+3332 & 0.890 & 186.743 & 33.547 & 33.089 & 81.693 &  &  &  & J \\
MACS J0647+7015 & 0.584 & 101.959 & 70.249 & 47.054 & 25.115 &  &  &  & J \\
MACS J0717.5+3745 & 0.548 & 109.383 & 37.753 & 15.370 & 21.046 &  &  &  & J \\
MACS J0744.8+3927 & 0.686 & 116.220 & 39.458 & 17.911 & 26.652 &  &  &  & J \\
MACS J1149+2223 & 0.544 & 177.395 & 22.401 & 19.447 & 75.197 &  &  &  & J \\
MACS J1423.8+2404 & 0.545 & 215.950 & 24.078 & 35.982 & 68.985 &  &  &  & J \\
MACS J1720.2+3536 & 0.391 & 260.070 & 35.607 & 58.511 & 33.077 & Y &  &  & J \\
MACS J2129-0741 & 0.589 & 322.359 & -7.691 & 6.774 & -38.465 &  & Y & Y & J \\
MS 2137.3-2353 & 0.313 & 325.063 & -23.661 & -9.182 & -46.937 &  & Y &  & J \\
RX J2129+0005 & 0.234 & 322.416 & 0.089 & 14.125 & -34.477 &  & Y & Y & J \\
MACS J0429.6-0253 & 0.399 & 67.400 & -2.885 & -24.390 & -32.588 &  & Y &  & K \\
MACS J1115.9+0129 & 0.352 & 168.966 & 1.499 & -2.987 & 55.625 &  & Y &  & K \\
MACS J1311.0-0310 & 0.494 & 197.757 & -3.178 & 4.030 & 59.328 &  & Y &  & K \\
MACS J1931.8-2634 & 0.352 & 292.957 & -26.576 & -4.755 & -20.093 &  & Y & Y & K \\
Abell 1489 & 0.350 & 183.079 & 27.553 & 26.318 & 81.304 &  &  &  & \cite{2020ApJ...903..137Z} \\ 
Abell 1689 & 0.187 & 197.873 & -1.3411 & 35.291 & 61.124 & & Y & & \cite{2007ApJ...668..643L} \\
PLCK G165.7+67.0 & 0.351 & 171.810 & 42.474 & 35.291 & 67.006 &  &  &  &  \cite{2024ApJ...961..171F} \\
\hline

\multicolumn{10}{l}{
\parbox{0.88\linewidth}{
\tablefoot{We list equatorial coordinates, ecliptic ($\beta$) and galactic ($b$) latitudes, as well as two conditions: whether the cluster has an ecliptic latitude $|\beta| > 54^\circ$, which is one of the HLTDS requirements, and whether the cluster falls within LSST observing fields.
A~-~The cluster was studied and modeled by the RELICS program \citep{2018ApJ...859..159C,2020ApJ...889..189S}, but no paper has been published on the cluster. 
B~-~The cluster can be observed by the HLTDS if the condition is relaxed to $|\beta| > 50^\circ$.
C~-~\cite{2018ApJ...859..159C}. 
D~-~\cite{2018ApJ...863..145C}.
E~-~\cite{2018ApJ...858...42A}.
F~-~the two parts of the cluster were modeled separately. 
G~-~\cite{2021AnA...646A..83R}.
H~-~\cite{2023AnA...678A.139T}.
I~-~The cluster was studied and modeled by \cite{2021AnA...646A..83R} and \cite{2023AnA...678A.139T}, but was not included in the papers. Data available at: \url{https://cral-perso.univ-lyon1.fr/labo/perso/johan.richard/MUSE_data_release/}, date of access: 2024-04-16.
J~-~\cite{2015ApJ...801...44Z}.
K~-~\cite{2019AnA...632A..36C}.
}
}
}
\end{longtable}

\begin{table}
\centering
\caption{Expected discovered multiply imaged SNe of given subtypes in \textit{Roman}, per cluster, depending on whether the cluster falls in the Deep or Wide survey tier, estimated with the volumetric method.}
\label{tab:vol_roman_percl}

\begin{tabular}{lcccccccc}
\hline
\hline
Cluster & Type IIP & Type IIL & Type IIb & Type IIn & Type Ia & Type Ib & Type Ic & Type Ic-BL \\ \hline
\multicolumn{9}{l}{Deep survey field}
\\ \hline
RXC J0232.2-4420 & 0.337$\pm$0.028 & 0.0546$\pm$0.0044 & 0.0655$\pm$0.0056 & 0.0334$\pm$0.0027 & 0.246$\pm$0.012 & 0.0687$\pm$0.0058 & 0.0491$\pm$0.0041 & 0.00718$\pm$0.00060 \\ 
Abell S295 & 0.249$\pm$0.020 & 0.0417$\pm$0.0033 & 0.0483$\pm$0.0041 & 0.0258$\pm$0.0020 & 0.1994$\pm$0.0099 & 0.0513$\pm$0.0043 & 0.0369$\pm$0.0031 & 0.00540$\pm$0.00045 \\ 
Abell 1758a & 0.873$\pm$0.073 & 0.153$\pm$0.012 & 0.173$\pm$0.015 & 0.0950$\pm$0.0076 & 0.700$\pm$0.035 & 0.187$\pm$0.016 & 0.136$\pm$0.012 & 0.0198$\pm$0.0017 \\ 
SMACS J0723.3-7327 & 0.273$\pm$0.023 & 0.0465$\pm$0.0038 & 0.0531$\pm$0.0046 & 0.0289$\pm$0.0023 & 0.225$\pm$0.011 & 0.0570$\pm$0.0049 & 0.0411$\pm$0.0035 & 0.00602$\pm$0.00051 \\ 
Bullet & 0.470$\pm$0.040 & 0.0812$\pm$0.0067 & 0.0922$\pm$0.0081 & 0.0505$\pm$0.0041 & 0.383$\pm$0.019 & 0.0991$\pm$0.0086 & 0.0717$\pm$0.0062 & 0.01048$\pm$0.00089 \\ 
SPT-CLJ0615-5746 & 0.602$\pm$0.055 & 0.121$\pm$0.011 & 0.115$\pm$0.011 & 0.0786$\pm$0.0068 & 0.714$\pm$0.040 & 0.134$\pm$0.013 & 0.0995$\pm$0.0095 & 0.0147$\pm$0.0014 \\ 
Abell 2261 & 0.572$\pm$0.049 & 0.1026$\pm$0.0085 & 0.114$\pm$0.010 & 0.0642$\pm$0.0052 & 0.466$\pm$0.024 & 0.125$\pm$0.011 & 0.0915$\pm$0.0079 & 0.0133$\pm$0.0011 \\ 
MACS J0553.4-3342 & 1.81$\pm$0.16 & 0.347$\pm$0.030 & 0.359$\pm$0.033 & 0.223$\pm$0.019 & 1.699$\pm$0.089 & 0.412$\pm$0.037 & 0.305$\pm$0.028 & 0.0443$\pm$0.0039 \\ 
MACS J1720.2+3536 & 0.319$\pm$0.027 & 0.0566$\pm$0.0046 & 0.0624$\pm$0.0055 & 0.0354$\pm$0.0028 & 0.280$\pm$0.014 & 0.0678$\pm$0.0059 & 0.0493$\pm$0.0043 & 0.00722$\pm$0.00062 \\ \hline
\multicolumn{9}{l}{Wide survey field}
\\ \hline
RXC J0232.2-4420 & 0.211$\pm$0.018 & 0.0395$\pm$0.0034 & 0.0360$\pm$0.0035 & 0.0266$\pm$0.0022 & 0.1425$\pm$0.0087 & 0.0404$\pm$0.0038 & 0.0297$\pm$0.0028 & 0.00454$\pm$0.00042 \\ 
Abell S295 & 0.147$\pm$0.013 & 0.0285$\pm$0.0024 & 0.0243$\pm$0.0023 & 0.0197$\pm$0.0016 & 0.1049$\pm$0.0064 & 0.0278$\pm$0.0026 & 0.0206$\pm$0.0020 & 0.00318$\pm$0.00029 \\ 
Abell 1758a & 0.472$\pm$0.042 & 0.0983$\pm$0.0085 & 0.0795$\pm$0.0077 & 0.0701$\pm$0.0058 & 0.382$\pm$0.024 & 0.0958$\pm$0.0092 & 0.0727$\pm$0.0070 & 0.0112$\pm$0.0010 \\ 
SMACS J0723.3-7327 & 0.160$\pm$0.014 & 0.0311$\pm$0.0027 & 0.0260$\pm$0.0025 & 0.0218$\pm$0.0018 & 0.1148$\pm$0.0072 & 0.0301$\pm$0.0029 & 0.0224$\pm$0.0022 & 0.00346$\pm$0.00032 \\ 
Bullet & 0.260$\pm$0.023 & 0.0529$\pm$0.0046 & 0.0435$\pm$0.0043 & 0.0376$\pm$0.0032 & 0.200$\pm$0.013 & 0.0511$\pm$0.0050 & 0.0383$\pm$0.0037 & 0.00592$\pm$0.00056 \\ 
SPT-CLJ0615-5746 & 0.245$\pm$0.024 & 0.0604$\pm$0.0059 & 0.0327$\pm$0.0038 & 0.0497$\pm$0.0046 & 0.247$\pm$0.018 & 0.0443$\pm$0.0051 & 0.0346$\pm$0.0040 & 0.00581$\pm$0.00061 \\ 
Abell 2261 & 0.303$\pm$0.027 & 0.0640$\pm$0.0056 & 0.0514$\pm$0.0050 & 0.0464$\pm$0.0039 & 0.254$\pm$0.016 & 0.0630$\pm$0.0061 & 0.0483$\pm$0.0047 & 0.00739$\pm$0.00069 \\ 
MACS J0553.4-3342 & 0.868$\pm$0.080 & 0.194$\pm$0.018 & 0.139$\pm$0.014 & 0.150$\pm$0.013 & 0.792$\pm$0.052 & 0.180$\pm$0.018 & 0.139$\pm$0.014 & 0.0217$\pm$0.0021 \\ 
MACS J1720.2+3536 & 0.165$\pm$0.015 & 0.0351$\pm$0.0031 & 0.0265$\pm$0.0026 & 0.0255$\pm$0.0021 & 0.1343$\pm$0.0085 & 0.0319$\pm$0.0031 & 0.0242$\pm$0.0024 & 0.00379$\pm$0.00036 \\  \hline

\end{tabular}

\end{table}

\begin{table}
\small
\centering
\caption{Expected discovered multiply imaged SNe of given subtypes in LSST, per cluster, estimates from the volumetric method.}
\label{tab:vol_lsst_percl}

\begin{tabular}{lccccccc}
\hline
\hline
Cluster & Type Ia & Type IIP+IIL & Type IIb & Type IIn & Type Ib & Type Ic & Type Ic-BL \\ \hline
Abell 209 & 0.0109$\pm$0.0033 & 0.0133$\pm$0.0053 & 0.00093$\pm$0.00036 & 0.0039$\pm$0.0016 & 0.00150$\pm$0.00058 & 0.00123$\pm$0.00048 & 0.000205$\pm$0.000081 \\ 
MACS J0257.6-2209 & 0.038$\pm$0.012 & 0.059$\pm$0.024 & 0.0040$\pm$0.0016 & 0.0143$\pm$0.0060 & 0.0061$\pm$0.0024 & 0.0050$\pm$0.0020 & 0.00079$\pm$0.00032 \\ 
MACS J0257.1-2325 & 0.0232$\pm$0.0070 & 0.031$\pm$0.013 & 0.00152$\pm$0.00060 & 0.0126$\pm$0.0053 & 0.0028$\pm$0.0011 & 0.00235$\pm$0.00093 & 0.00048$\pm$0.00020 \\ 
MACS J0329.6-0211 & 0.066$\pm$0.020 & 0.106$\pm$0.044 & 0.0059$\pm$0.0023 & 0.029$\pm$0.012 & 0.0095$\pm$0.0038 & 0.0078$\pm$0.0031 & 0.00140$\pm$0.00057 \\ 
MACS J0429.6-0253 & 0.0122$\pm$0.0037 & 0.0144$\pm$0.0058 & 0.00080$\pm$0.00031 & 0.0058$\pm$0.0024 & 0.00142$\pm$0.00056 & 0.00123$\pm$0.00049 & 0.000237$\pm$0.000096 \\ 
Abell 2537 & 0.0287$\pm$0.0087 & 0.042$\pm$0.017 & 0.00241$\pm$0.00095 & 0.0117$\pm$0.0050 & 0.0040$\pm$0.0016 & 0.0032$\pm$0.0013 & 0.00057$\pm$0.00023 \\ 
MACS J0416.1-2403 & 0.045$\pm$0.014 & 0.060$\pm$0.025 & 0.0031$\pm$0.0012 & 0.023$\pm$0.010 & 0.0055$\pm$0.0022 & 0.0047$\pm$0.0019 & 0.00090$\pm$0.00037 \\ 
MACS J1115.9+0129 & 0.00258$\pm$0.00079 & 0.0058$\pm$0.0026 & 0.000147$\pm$0.000062 & 0.00170$\pm$0.00076 & 0.00025$\pm$0.00011 & 0.000231$\pm$0.000097 & (5.7$\pm$2.5)e-05 \\ 
RXC J2211.7-0350 & 0.080$\pm$0.024 & 0.084$\pm$0.034 & 0.0046$\pm$0.0018 & 0.040$\pm$0.017 & 0.0088$\pm$0.0034 & 0.0076$\pm$0.0030 & 0.00150$\pm$0.00061 \\ 
Abell 2744 & 0.073$\pm$0.022 & 0.092$\pm$0.038 & 0.0056$\pm$0.0022 & 0.031$\pm$0.013 & 0.0094$\pm$0.0037 & 0.0078$\pm$0.0031 & 0.00142$\pm$0.00057 \\ 
MACS J0417.5-1154 & 0.040$\pm$0.012 & 0.053$\pm$0.022 & 0.0026$\pm$0.0010 & 0.0219$\pm$0.0094 & 0.0046$\pm$0.0018 & 0.0040$\pm$0.0016 & 0.00087$\pm$0.00036 \\ 
MACS J1311.0-0310 & 0.0084$\pm$0.0025 & 0.0129$\pm$0.0053 & 0.00065$\pm$0.00026 & 0.0046$\pm$0.0020 & 0.00113$\pm$0.00045 & 0.00092$\pm$0.00037 & 0.000176$\pm$0.000072 \\ 
RX J1347.5-1145 & 0.038$\pm$0.012 & 0.050$\pm$0.021 & 0.0027$\pm$0.0011 & 0.0206$\pm$0.0088 & 0.0047$\pm$0.0018 & 0.0039$\pm$0.0015 & 0.00083$\pm$0.00034 \\ 
Abell 2813 & 0.0188$\pm$0.0057 & 0.027$\pm$0.011 & 0.00134$\pm$0.00053 & 0.0093$\pm$0.0040 & 0.00238$\pm$0.00095 & 0.00200$\pm$0.00080 & 0.00038$\pm$0.00015 \\ 
MACS J0451.9+0006 & 0.0186$\pm$0.0057 & 0.033$\pm$0.014 & 0.00130$\pm$0.00053 & 0.0113$\pm$0.0050 & 0.00223$\pm$0.00091 & 0.00192$\pm$0.00079 & 0.00043$\pm$0.00018 \\ 
MACS J1931.8-2634 & 0.0232$\pm$0.0070 & 0.025$\pm$0.010 & 0.00134$\pm$0.00053 & 0.0105$\pm$0.0044 & 0.0026$\pm$0.0010 & 0.00230$\pm$0.00090 & 0.00042$\pm$0.00017 \\ 
RX J2129+0005 & 0.0107$\pm$0.0032 & 0.0153$\pm$0.0063 & 0.00095$\pm$0.00037 & 0.0042$\pm$0.0018 & 0.00150$\pm$0.00059 & 0.00123$\pm$0.00049 & 0.000213$\pm$0.000086 \\ 
Abell 3192 & 0.089$\pm$0.027 & 0.131$\pm$0.054 & 0.0077$\pm$0.0030 & 0.034$\pm$0.014 & 0.0127$\pm$0.0051 & 0.0103$\pm$0.0041 & 0.00178$\pm$0.00072 \\ 
MACS J0520.7-1328 & 0.049$\pm$0.015 & 0.075$\pm$0.031 & 0.0049$\pm$0.0019 & 0.0181$\pm$0.0077 & 0.0076$\pm$0.0030 & 0.0061$\pm$0.0024 & 0.00105$\pm$0.00043 \\ 
MACS J2129-0741 & 0.0125$\pm$0.0038 & 0.0212$\pm$0.0091 & 0.00085$\pm$0.00035 & 0.0071$\pm$0.0031 & 0.00148$\pm$0.00061 & 0.00126$\pm$0.00052 & 0.00026$\pm$0.00011 \\ 
SMACS J2031.8-4036 & 0.0159$\pm$0.0048 & 0.025$\pm$0.010 & 0.00125$\pm$0.00049 & 0.0078$\pm$0.0034 & 0.00209$\pm$0.00083 & 0.00171$\pm$0.00068 & 0.00034$\pm$0.00014 \\ 
Abell 370 & 0.079$\pm$0.024 & 0.116$\pm$0.048 & 0.0063$\pm$0.0025 & 0.033$\pm$0.014 & 0.0107$\pm$0.0042 & 0.0088$\pm$0.0035 & 0.00160$\pm$0.00065 \\ 
MACS J0553.4-3342 & 0.080$\pm$0.024 & 0.128$\pm$0.053 & 0.0068$\pm$0.0027 & 0.042$\pm$0.018 & 0.0107$\pm$0.0043 & 0.0088$\pm$0.0036 & 0.00188$\pm$0.00078 \\ 
MS 1008.1-1224 & 0.0254$\pm$0.0076 & 0.030$\pm$0.012 & 0.00174$\pm$0.00068 & 0.0114$\pm$0.0048 & 0.0030$\pm$0.0012 & 0.0026$\pm$0.0010 & 0.00051$\pm$0.00021 \\ 
SMACS J2131.1-4019 & 0.0243$\pm$0.0073 & 0.036$\pm$0.015 & 0.00217$\pm$0.00086 & 0.0099$\pm$0.0042 & 0.0036$\pm$0.0014 & 0.0029$\pm$0.0011 & 0.00051$\pm$0.00021 \\ 
Abell 383 & 0.0283$\pm$0.0085 & 0.034$\pm$0.014 & 0.00257$\pm$0.00099 & 0.0091$\pm$0.0038 & 0.0041$\pm$0.0016 & 0.0033$\pm$0.0013 & 0.00054$\pm$0.00021 \\ 
MACS J0940.9+0744 & 0.0122$\pm$0.0037 & 0.0147$\pm$0.0059 & 0.00096$\pm$0.00037 & 0.0048$\pm$0.0020 & 0.00159$\pm$0.00062 & 0.00132$\pm$0.00052 & 0.000242$\pm$0.000097 \\ 
MS 2137.3-2353 & 0.0126$\pm$0.0038 & 0.0171$\pm$0.0070 & 0.00098$\pm$0.00038 & 0.0058$\pm$0.0025 & 0.00166$\pm$0.00066 & 0.00140$\pm$0.00055 & 0.00025$\pm$0.00010 \\ 
SMACS J2332.4-5358 & 0.0139$\pm$0.0042 & 0.0190$\pm$0.0078 & 0.00093$\pm$0.00037 & 0.0077$\pm$0.0033 & 0.00169$\pm$0.00067 & 0.00143$\pm$0.00057 & 0.00029$\pm$0.00012 \\ 
ACT-CLJ0102-49151 & 0.00213$\pm$0.00066 & 0.0077$\pm$0.0035 & (6.5$\pm$2.9)e-05 & 0.0048$\pm$0.0022 & 0.000111$\pm$0.000049 & 0.000115$\pm$0.000051 & 0.000107$\pm$0.000050 \\ 
MACS J1206.2-0847 & 0.041$\pm$0.012 & 0.060$\pm$0.025 & 0.0032$\pm$0.0013 & 0.0180$\pm$0.0077 & 0.0053$\pm$0.0021 & 0.0045$\pm$0.0018 & 0.00084$\pm$0.00035 \\ 
PLCK G209.79+10.23 & 0.0123$\pm$0.0037 & 0.0226$\pm$0.0097 & 0.00066$\pm$0.00027 & 0.0105$\pm$0.0046 & 0.00120$\pm$0.00050 & 0.00105$\pm$0.00044 & 0.00033$\pm$0.00014 \\ 
Abell S1063 & 0.0218$\pm$0.0066 & 0.028$\pm$0.011 & 0.00197$\pm$0.00077 & 0.0073$\pm$0.0030 & 0.0031$\pm$0.0012 & 0.00254$\pm$0.00099 & 0.00041$\pm$0.00016 \\ 
MACS J2214.9-1359 & 0.0176$\pm$0.0053 & 0.029$\pm$0.012 & 0.00125$\pm$0.00050 & 0.0094$\pm$0.0041 & 0.00216$\pm$0.00087 & 0.00179$\pm$0.00072 & 0.00037$\pm$0.00015 \\ 
PLCK G287.0+32.9 & 0.077$\pm$0.023 & 0.113$\pm$0.047 & 0.0063$\pm$0.0025 & 0.033$\pm$0.014 & 0.0108$\pm$0.0043 & 0.0085$\pm$0.0034 & 0.00150$\pm$0.00061 \\ 
SPT-CLJ0615-5746 & 0.0060$\pm$0.0019 & 0.0150$\pm$0.0067 & 0.000213$\pm$0.000093 & 0.0094$\pm$0.0043 & 0.00041$\pm$0.00018 & 0.00041$\pm$0.00018 & 0.000207$\pm$0.000094 \\ 
Abell S295 & 0.0180$\pm$0.0054 & 0.028$\pm$0.011 & 0.00153$\pm$0.00061 & 0.0076$\pm$0.0032 & 0.0025$\pm$0.0010 & 0.00207$\pm$0.00082 & 0.00035$\pm$0.00014 \\ 
MACS J0025.4-1222 & 0.045$\pm$0.014 & 0.053$\pm$0.022 & 0.00212$\pm$0.00085 & 0.037$\pm$0.016 & 0.0043$\pm$0.0017 & 0.0040$\pm$0.0016 & 0.00101$\pm$0.00042 \\ 
WHL J013719.8-082841 & 0.0183$\pm$0.0055 & 0.027$\pm$0.011 & 0.00101$\pm$0.00041 & 0.0128$\pm$0.0056 & 0.00195$\pm$0.00079 & 0.00170$\pm$0.00069 & 0.00040$\pm$0.00017 \\ 
Bullet & 0.0311$\pm$0.0094 & 0.044$\pm$0.018 & 0.0027$\pm$0.0011 & 0.0127$\pm$0.0054 & 0.0045$\pm$0.0018 & 0.0037$\pm$0.0015 & 0.00062$\pm$0.00025 \\ 
MACS J0035.4-2015 & 0.050$\pm$0.015 & 0.064$\pm$0.026 & 0.0039$\pm$0.0015 & 0.0208$\pm$0.0087 & 0.0066$\pm$0.0026 & 0.0055$\pm$0.0022 & 0.00094$\pm$0.00038 \\ 
RXC J0232.2-4420 & 0.035$\pm$0.011 & 0.057$\pm$0.023 & 0.0036$\pm$0.0014 & 0.0131$\pm$0.0056 & 0.0055$\pm$0.0022 & 0.0044$\pm$0.0017 & 0.00074$\pm$0.00030 \\ 
CLJ0152.7-1357 & 0.00141$\pm$0.00043 & 0.0033$\pm$0.0015 & (7.7$\pm$3.3)e-05 & 0.00166$\pm$0.00076 & 0.000121$\pm$0.000052 & 0.000118$\pm$0.000050 & (4.5$\pm$2.0)e-05 \\ 
MACS J0159.8-0849 & 0.043$\pm$0.013 & 0.067$\pm$0.028 & 0.0037$\pm$0.0015 & 0.0194$\pm$0.0083 & 0.0059$\pm$0.0024 & 0.0049$\pm$0.0020 & 0.00090$\pm$0.00037 \\ 
RXC J0600.1-2007 & 0.0266$\pm$0.0081 & 0.047$\pm$0.020 & 0.00191$\pm$0.00077 & 0.0144$\pm$0.0063 & 0.0032$\pm$0.0013 & 0.0027$\pm$0.0011 & 0.00055$\pm$0.00023 \\
Abell 1689 & 0.1007$\pm$0.0091 & 0.142$\pm$0.040 & 0.0091$\pm$0.0024 & 0.038$\pm$0.011 & 0.0146$\pm$0.0039 & 0.0118$\pm$0.0032 & 0.00195$\pm$0.00054 \\
 \hline

\end{tabular}

\end{table}

\end{landscape}
\end{document}